\documentclass[aps,pra,preprint,showpacs,amsmath,amssymb]{revtex4}
\input{epsf}

\usepackage{graphicx}
\usepackage{array}
\usepackage{bigstrut}
\usepackage{longtable}
\usepackage{rotating,booktabs}
\usepackage{booktabs,threeparttable}
\usepackage{bm}

\begin{document}

\title{\bf
High-precision nonadiabatic calculations of dynamic polarizabilities
and hyperpolarizabilities for the lowlying vibrational-rotational
states of hydrogen molecular ions}
\author{Li-Yan Tang$^{1,3}$, Zong-Chao Yan$^{1,2,3}$, Ting-Yun Shi$^{1}$,
and James F. Babb$^3$}

\affiliation {$^1$State Key Laboratory of Magnetic Resonance and
Atomic and Molecular Physics and Center for Cold Atom Physics, Wuhan
Institute of Physics and Mathematics, Chinese Academy of Sciences,
Wuhan 430071, P. R. China}

\affiliation{$^2$Department of Physics, University of New
Brunswick, Fredericton, New Brunswick, Canada E3B 5A3}

\affiliation {$^{3}$ITAMP, Harvard-Smithsonian Center for
Astrophysics, 60 Garden Street, Cambridge, Massachusetts 02138, USA}

\date{\today}

\begin{abstract}
{The static and dynamic electric multipolar polarizabilities and
second hyperpolarizabilities of the H$_2^+$, D$_2^+$, and HD$^+$
molecular ions in the ground and first excited states are calculated
nonrelativistically using explicitly correlated Hylleraas basis
sets. The calculations are fully nonadiabatic; the Born-Oppenheimer
approximation is not used. Comparisons are made with published
theoretical and experimental results, where available. In our
approach, no derivatives of energy functions nor derivatives of
response functions are needed. In particular, we make contact with
earlier calculations in the Born-Oppenheimer calculation where
polarizabilities were decomposed into electronic, vibrational, and
rotational contributions and where hyperpolarizabilities were
determined from derivatives of energy functions. We find that the
static hyperpolarizability for the ground state of HD$^+$ is seven
orders of magnitude larger than the corresponding dipole
polarizability. For the dipole polarizability of HD$^+$ in the first
excited-state the high precision of the present method facilitates
treatment of a near cancellation between two terms. For applications
to laser spectroscopy of trapped ions we find tune-out and magic
wavelengths for the HD$^+$ ion in a laser field. In addition, we
also calculate the first few leading terms for long-range
interactions of a hydrogen molecular ion and a ground-state H, He,
or Li atom.}
\end{abstract}

 \maketitle

\section{Introduction}
Polarizabilities and hyperpolarizabilities of molecules can describe
linear and nonlinear optical phenomena, such as light scattering
from gases and solids and the Kerr effect, and dynamic (or
frequency-dependent) values are helpful in designing optical
materials and in gauging electric field responses for experiments.
While calculations are challenging, there are numerous calculated
results for many molecules---static and dynamic polarizabilities and
hyperpolarizabilities are available properties in many mature
quantum chemistry programs---yet actual fully nonadiabatic ab initio
results (obtained without use of the Born-Oppenheimer picture) are
rare. In previous studies, it was
demonstrated~\cite{yan03b,zhang04a,bhatia99a,bhatia00a} that a
theory based on the explicitly correlated Hylleraas basis set
expansion yielded high accuracy nonadiabatic properties of
three-body systems. In this paper, we extend the formalism
contiguously to multipolar dynamic electric polarizabilities and
dynamic second hyperpolarizabilities of the hydrogen molecular ion
and its deuterium containing isotopologues in the ground and first
excited states. While the formalism presented here is purely
nonrelativisitic, the nonadiabatic theory on which it is based is
well-tested beyond order $\alpha^2~\textrm{Ry}$ as progress in
calculations of energies of $\textrm{HD}^+$, for example, are now at
the level where the uncertainties in transition frequencies are of
the order of 70~kHz, with unknown effects contributing at order
$\alpha^5~\textrm{Ry}$~\cite{ZhoZhaYan12}, while refinement of
nonadiabatic calculations on simple molecules continues using
different
approaches~\cite{OliPilBay13,StaAda13,KorZho12,KedStaBub06}. A
comparison of nonrelativistic results for energies is given in
Sec.~\ref{results}.

The present calculations, we believe, are of great value for several
potential applications. While our approach intrinsically includes
rotational and vibrational degrees of freedom it dispenses with the
Born-Oppenheimer approximation. Use of the Born-Oppenheimer
approximation facilitates the breakdown of polarizabilities and
hyperpolarizabilities into ``electronic'', ``vibrational'', and
``rotational'' components and the theoretical underpinnings of this
picture are well-established, but there are different formulations
and subtleties in executing such
calculations~\cite{BisCheBuc80,PanSan80,AdaBar86,Bis90,BisKir92,SheRic94,Bis98}.
We show how our results provide insight into these descriptions,
allowing direct comparisons with earlier Born-Oppenheimer results,
and in Sec.~\ref{results} we use these insights, for example, to
resolve a discrepancy found by Olivares Pil\'on and
Baye~\cite{OliPilBay12} in comparing nonadiabatic and
Born-Oppenheimer calculations of the dynamic electric quadrupole
polarizability. Our method avoids the cumbersome Born-Oppenheimer
separation, our tabulated nonadiabatic data can be valuable for
estimations or extrapolations of ``electronic", ``vibrational", and
``rotational" contributions, when combined with available
Born-Oppenheimer calculations~\cite{SheRic94,koelemeij11a}. In
addition, our nonadiabatic approach does not require derivatives of
an energy function~\cite{PanSan80,CafBubAda03} nor derivatives of
response functions~\cite{IngPapHan98}, which can introduce
additional numerical loss of precision, but it does provide
definitive convergence-based error bars thereby allowing us to gauge
the accuracy of previous results for hyperpolarizabilities
calculated using gradients of fields.

There is much recent interest in trapping molecular ions for precision measurements
(of time~\cite{SchBakKor14} and of mass~\cite{KarHilKor11}, for example) and for
realizing quantum computing~\cite{ShiHerDre13}---in these
cases the responses of ions to applied fields are important considerations~\cite{BakSch12} and our calculations
can serve as useful models or references for future studies.
We find, for example, that the hyperpolarizabilities of
H$_2^+$ and D$_2^+$ are much larger than the dipole polarizabilities by four orders
of magnitude, which confirms~\cite{BisSol85} that the Stark shift of H$_2^+$ immersed
at high field strength would be influenced by the
hyperpolarizability.
For the ground state of HD$^+$, the sign of static dipole polarizability and hyperpolarizability are opposite,
suggesting that the hyperpolarizability should be considered in experimental analyses, since the Stark shifts for this system would tend to cancel each other.
In Sec.~\ref{results} we present highly accurate calculations of Stark shifts,
tune out and magic wavelengths, and
nonlinear dynamic hyperpolarizabilities for $\textrm{HD}^+$
in the ground and excited states.
Finally, the multipolar polarizabilities that we compute enter as parameters  in the long-range
``polarization potential''~\cite{sturrus91a,jacobson97a,jacobson00a}, which are effective
potential
expansions, for the interactions of an electron with the
the molecular ion isotopologues. We also calculate the long-range dispersion interactions
between H, He, or Li and each of the $\textrm{H}_2^+$ isotopologues in their ground
or first excited states.

In this work, the 2006 CODATA masses~\cite{mohr08a} of the proton
and the deuteron are adopted~\cite{footnote},
where
\begin{eqnarray}
m_p&=&1836.152 672 47(80)m_e, \\
m_d&=&3670.482 9654(16)m_e,
\end{eqnarray}
and $m_e$ is the electron mass, and atomic units are used throughout
unless specifically mentioned. The polarizabilities and
hyperpolarizabilities are presented in atomic units~\cite{SheRic94};
conversion factors to SI units are given in, for example, the
reviews by Bishop~\cite{Bis90} and by Shelton and
Rice~\cite{SheRic94}. In this nonrelativistic study we neglect
finite temperature effects~\cite{BisLamEps88,SheRic94}, hyperfine
structure~\cite{SchBakKor14,SchBakKor14b}, and we do not consider
the first hyperpolarizability (which is only non zero for
$\textrm{HD}^+$).

\section{THEORY AND METHOD}

\subsection{Hamiltonian and Hylleraas basis}

In the present work, we treat the hydrogen molecular ion as a
three-body Coulombic system; the calculations are fully nonadiabatic
(the Born-Oppenheimer approximation is not used). Taking one of the
nuclei as particle 0, the electron is chosen as particle 1 and the
other nucleus is seen as particle 2. In the center of mass frame,
the Hamiltonian can be written as
\begin{eqnarray}
H_0 &=& -\sum_{i=1}^2 \frac{1}{2\mu_i}\nabla_i^2
-\frac{1}{m_0}\sum_{i> j\ge 1}^2\nabla_i\cdot\nabla_j
+q_0\sum_{i=1}^2\frac{q_i}{r_i} +\sum_{i> j\ge
1}^2\frac{q_iq_j}{r_{ij}}\,, \label {eq:h0}
\end{eqnarray}
where $\mu_i=m_im_0/(m_i+m_0)$ is the reduced mass between particle
$i$ and particle $0$, $q_i$ is the charge of the $i$th particle,
$\mathbf{r}_i$ is the position vector between particle $i$ and particle $0$,
and $r_{ij}=|{\bf{r}}_{i}-{\bf{r}}_{j}|$ is the inter-particle
separation.

The wave functions are constructed in terms of the explicitly
correlated Hylleraas coordinates as
\begin{eqnarray}
\phi_{ijk}({\bf{r}}_{1},{\bf{r}}_{2})=r_{1}^{i}r_{2}^{j}r_{12}^{k}e^{-\alpha
r_{1}-\beta
r_{2}}\mathcal{Y}_{\ell_{1}\ell_{2}}^{LM}(\hat{{\bf{r}}}_{1},\hat{{\bf{r}}}_{2}),
\label {eq:t18}
\end{eqnarray}
where $r_2^je^{-\beta r_2}$ sufficiently represents the vibrational
modes between the nuclei if $j$ and $\beta$ are chosen big
enough~\cite{yan03b},
$\mathcal{Y}_{\ell_{1}\ell_{2}}^{LM}(\hat{{\bf{r}}}_{1},\hat{{\bf{r}}}_{2})$
is a vector-coupled product of spherical harmonics,
\begin{eqnarray}
&&\mathcal{Y}_{\ell_{1}\ell_{2}}^{LM}(\hat{{\bf{r}}}_{1},
\hat{{\bf{r}}}_{2}) =\sum_{ m_1,m_2}\langle
\ell_{1}m_{1};\ell_{2}m_{2}|LM\rangle
Y_{\ell_{1}m_{1}}(\hat{{\bf{r}}}_{1})
Y_{\ell_{2}m_{2}}(\hat{{\bf{r}}}_{2})\,, \label {eq:t19}
\end{eqnarray}
and the nonlinear parameters $\alpha$ and $\beta$ are optimized using Newton's method. All terms in Eq.(\ref{eq:t18}) are included such that
\begin{eqnarray}
i+j+k\leq \Omega\,,
\end{eqnarray}
where $\Omega$ is an integer, and the convergence for the energy
eigenvalue is studied as $\Omega$ is increased progressively. The
computational details used in evaluating the necessary matrix
elements of the Hamiltonian are given in~Ref.~\cite{YanDra96}.

\subsection{Polarizability and Hyperpolarizability}

When the hydrogen molecular ion is exposed to a weak external electric field ${\cal{E}}$, the
second-order Stark shift for the rovibronic state is
\begin{eqnarray} \Delta
E_2=-\frac{{\cal{E}}^2}{2}[\alpha_1(\omega)+\alpha_1^{(T)}(\omega)g_2(L,M)]\,,\label
{eq:t5}
\end{eqnarray}
where $L$ is the angular momentum with magnetic quantum number $M$,
$g_2(L,M)$ is the only $M$-dependent part,
\begin{eqnarray}
g_2(L,M)&=&\frac{3M^2-L(L+1)}{L(2L-1)}, \ \ L\ge 1\,, \label {eq:t6}
\end{eqnarray}
and $\omega$ is the frequency  of the external electric field in the
$z$-direction. The dynamic scalar and tensor dipole
polarizabilities, respectively, are $\alpha_1(\omega)$ and
$\alpha_1^{(T)}(\omega)$; when $\omega=0$, they are called,
respectively, the static scalar and tensor dipole polarizabilities.
The derivation of  the expressions for the dynamic polarizabilities
$\alpha_1(\omega)$ and $\alpha_1^{(T)}(\omega)$ are similar to those
described in Ref.~\cite{tang10a}. In particular, for the case of
rovibronic ground-state with $L=0$,
\begin{eqnarray}
\alpha_1(\omega)=\alpha_1(P,\omega)\,, \qquad \alpha_1^{(T)}(\omega)=0 ,
\end{eqnarray}
with $\alpha_1(L_a,\omega)$ following the general expression of $2^\ell$-pole partial dynamic polarizabilities,
\begin{eqnarray}
\alpha_\ell(L_a,\omega) &=& \frac{8\pi}{(2\ell+1)^2(2L+1)}\sum_n \frac{\Delta E_{n0}|\langle n_0 L
\|T_\ell\| nL_a\rangle |^2}{\Delta E_{n0}^2-\omega^2}\,,
 \label {eq:t10}
\end{eqnarray}
where $n_0$ and $n$, respectively, label the  initial state and the
intermediate state and $\Delta E_{n0}=E_{n}-E_{n_0}$ is the
difference between the initial and intermediate state energies. The
detailed formula for the  $2^{\ell}$-pole transition operator
$T_\ell$ in the center of mass frame is given in
Ref.~\cite{tang09a}.

For the rovibronic excited-state with $L=1$, $\alpha_1(\omega)$ and $\alpha_1^{(T)}(\omega)$ can be written
\begin{eqnarray}
\label {eq:t26_1}
\alpha_1(\omega) &=& \alpha_1(S,\omega)+\alpha_1(P,\omega)+\alpha_1(D,\omega)\, , \label {eq:t8} \\
\alpha_1^{(T)}(\omega) &=&
-\alpha_1(S,\omega)+\frac{1}{2}\alpha_1(P,\omega)-\frac{1}{10}\alpha_1(D,\omega)\, , \label {eq:t9}
\end{eqnarray}
where $\alpha_1(P,\omega)$ denotes the
contribution of nucleus 2 and electron 1 both being in $p$ configuration to form a
total angular momentum of $P$. The expressions for other multipole dynamic polarizabilities are derived similarly to those for the dipole polarizabilities~\cite{tang09a,tang09b,tang10a}.

The fourth-order Stark shift for the rovibronic state can be written in the form,
\begin{eqnarray}
\Delta E_4 &=&
-\frac{{\cal{E}}^4}{24}\bigg[\gamma_0(-\omega_\sigma;\omega_1,\omega_2,\omega_3)+\gamma_2(-\omega_\sigma;\omega_1,\omega_2,\omega_3)\,g_2(L,M)
+\gamma_4(-\omega_\sigma;\omega_1,\omega_2,\omega_3)\,g_4(L,M)\bigg]\,, \nonumber\\
 \label {eq:t34}
\end{eqnarray}
where $g_4(L,M)$ is only dependent on the angular momentum quantum number $L$ and magnetic
quantum number $M$,
\begin{eqnarray}
g_4(L,M)= \frac{3(5M^2-L^2-2L)(5M^2+1-L^2)-10M^2(4M^2-1)}{L(2L-1)(2L-2)(2L-3)},  \, L \geq 2 ,
\label {eq:a38}
\end{eqnarray}
and $\omega_i$ are the frequencies of the external electric field in
the three directions with
$\omega_\sigma=\omega_1+\omega_2+\omega_3$. The dynamic scalar
second hyperpolarizability is
$\gamma_0(-\omega_\sigma;\omega_1,\omega_2,\omega_3)$, and the
dynamic tensor second hyperpolarizabilities are
$\gamma_2(-\omega_\sigma;\omega_1,\omega_2,\omega_3)$ and
$\gamma_4(-\omega_\sigma;\omega_1,\omega_2,\omega_3)$. (From this
point on, we will omit ``second" when referring to the
hyperpolarizabilities.) When all $\omega_i=0$, the functions are
called static hyperpolarizabilities. In particular, for the
rovibronic excited-state with $L=0$ only the dynamic scalar
hyperpolarizability remains and it is
\begin{eqnarray}
\gamma_0(-\omega_\sigma;\omega_1,\omega_2,\omega_3)=\frac{16\pi^2}{9}\big[\frac{1}{9}\mathcal{T}(1,0,1;\omega_1,\omega_2,\omega_3)
+\frac{2}{45}\mathcal{T}(1,2,1;\omega_1,\omega_2,\omega_3)\big] , \label {hyper}
\end{eqnarray}
where
\begin{eqnarray}
\mathcal{T}(L_a,L_b,L_c;\omega_1,\omega_2,\omega_3) &=& \sum_{P}\bigg[\sum_{kmn}\frac{\langle
n_0L\|T_1^{\mu_1}\|mL_a\rangle \langle mL_a\|T_1^{\mu_2}\|nL_b\rangle \langle
nL_b\|T_1^{\mu_3}\|kL_c\rangle \langle kL_c\|T_1^{\mu_4}\|n_0L\rangle }
{(\Delta E_{mn_0}-\omega_\sigma)(\Delta E_{nn_0}-\omega_1-\omega_2)(\Delta E_{kn_0}-\omega_1)}\nonumber
\\
&-& \delta(L_b,L)
\sum_{m}
\frac{\langle n_0L\|T_1^{\mu_1}\|mL_a\rangle \langle mL_a\|T_1^{\mu_2}\| n_0L\rangle}
{(\Delta E_{mn_0}-\omega_\sigma)}
 \nonumber \\
&\times& \sum_{k}
\frac{\langle n_0L\|T_1^{\mu_3}\|kL_c\rangle \langle kL_c\|T_1^{\mu_4}\|n_0L\rangle }
{(\Delta E_{kn_0}+\omega_2)(\Delta E_{kn_0}-\omega_1)}\bigg] ,
\label {eq:a35}
\end{eqnarray}
the $\sum_P$ implies a summation over the 24 terms generated by
permuting the pairs ($-\omega_\sigma/T_1^{\mu_1}$),
($\omega_1/T_1^{\mu_2}$), ($\omega_2/T_1^{\mu_3}$), and
($\omega_3/T_1^{\mu_4}$), where the superscripts $\mu_i$ are
introduced for the purpose of labeling the
permutations~\cite{pipin92a}.

\section{ Results And Discussion}\label {results}

\subsection{Energies}
\begingroup
\squeezetable
\begin{table}[th]
\caption{Energies (in a.u.) for the H$_2^+$, D$_2^+$, and HD$^+$ ions for $\upsilon \leq 3$ and $L \leq 3 $.
For each value of $(\upsilon, L)$ in the first column, the first row gives the present result
resulting from a single diagonalization of the lowest $\upsilon$ state for a given $L$.
Where a value of  $(\upsilon, L)$ has a second row (H$_2^+$ and  HD$^+$) the entry on the second row lists the result of Korobov~\cite{korobov06a},
for which each value is the result of a separate minimization
(see text for further discussion).
The number in parentheses represents the computational uncertainty in the last digit.}\label{energy}
\begin{ruledtabular}
\begin{tabular}{llll}
\multicolumn{1}{l}{($\upsilon,L$)} & \multicolumn{1}{c}{H$_2^+$}
& \multicolumn{1}{c}{D$_2^+$} & \multicolumn{1}{c}{HD$^+$} \\
\hline
(0,0)    &$-$0.597 139 063 079 392 297 758(4)     & $-$0.598 788 784 304 562 857 67(6)   & $-$0.597 897 968 608 954 700 9(1)\\
         &$-$0.597 139 063 079 39                 &                                      & $-$0.597 897 968 609 03\\
(1,0)    &$-$0.587 155 679 164 695 13(2)          & $-$0.591 603 121 831 520 71(3)       & $-$0.589 181 829 556 745 7(1)\\
         &$-$0.587 155 679 096 19                 &                                      & $-$0.589 181 829 556 96\\
(2,0)    &$-$0.577 751 904 547 41(7)              & $-$0.584 712 206 896 55(1)           & $-$0.580 903 700 218(1)\\
         &$-$0.577 751 904 415 08                 &                                      & $-$0.580 903 700 218 37\\
(3,0)    &$-$0.568 908 498 91(7)                  & $-$0.578 108 591 285 37(2)           & $-$0.573 050 546(1)\\
         &$-$0.568 908 498 730 86                 &                                      & $-$0.573 050 546 551 87\\
(0,1)    &$-$0.596 873 738 784 713 077 8(1)       & $-$0.598 654 873 192 605 311 3(3)    & $-$0.597 698 128 192 126 71(1)\\
         &$-$0.596 873 738 784 71                 &                                      & $-$0.597 698 128 192 21\\
(1,1)    &$-$0.586 904 320 919 191 59(5)          & $-$0.591 474 211 455 255 47(6)       & $-$0.588 991 111 991 818(4)\\
         &$-$0.586 904 320 919 19                 &                                      & $-$0.588 991 111 992 04\\
(2,1)    &$-$0.577 514 034 057 4(2)               & $-$0.584 588 169 503 82(3)           & $-$0.580 721 828 12(1)\\
         &$-$0.577 514 034 057 45                 &                                      & $-$0.580 721 828 120 93\\
(3,1)    &$-$0.568 683 708 2(2)                   & $-$0.577 989 311 81(2)               & $-$0.572 877 277(3)\\
         &$-$0.568 683 708 260 19                 &                                      & $-$0.572 877 277 094 21\\
(0,2)    &$-$0.596 345 205 489 114 7(2)           & $-$0.598 387 585 778 605 864(3)      & $-$0.597 299 643 351 683 2(1)\\
         &$-$0.596 345 205 489 39                 &                                      & $-$0.597 299 643 351 78\\
(1,2)    &$-$0.586 403 631 528 0(8)               & $-$0.591 216 909 547 769 2(4)        & $-$0.588 610 829 389 5(2)\\
         &$-$0.586 403 631 528 69                 &                                      & $-$0.588 610 829 389 79\\
(2,2)    &$-$0.577 040 237 1(6)                   & $-$0.584 340 598 262 86(3)           & $-$0.580 359 195 2(6)\\
         &$-$0.577 040 237 163 02                 &                                      & $-$0.580 359 195 199 88\\
(3,2)    &$-$0.568 235 98(5)                      & $-$0.577 751 241 74(1)               & $-$0.572 531 8(2)\\
         &$-$0.568 235 992 971 58                 &                                      & $-$0.572 531 810 325 97\\
(0,3)    &$-$0.595 557 638 980 309 2(7)           & $-$0.597 987 984 710 141(4)          & $-$0.596 704 882 761 75(3)\\
         &$-$0.595 557 638 980 31                 &                                      & $-$0.596 704 882 761 89\\
(1,3)    &$-$0.585 657 611 877(1)                 & $-$0.590 832 246 988(4)              & $-$0.588 043 264 163(2)\\
         &$-$0.585 657 611 877 66                 &                                      & $-$0.588 043 264 162 84\\
(2,3)    &$-$0.576 334 350 2(2)                   & $-$0.583 970 493 6(1)                & $-$0.579 818 002 1(1)\\
         &$-$0.576 334 350 219 63                 &                                      & $-$0.579 818 002 027 87\\
(3,3)    &$-$0.567 569 02(5)                      & $-$0.577 395 352 1(9)                & $-$0.572 016 269 2(4)\\
         &$-$0.567 569 034 833 51                 &                                      & $-$0.572 016 269 232 51\\
\end{tabular}
\end{ruledtabular}
\end{table}
\endgroup

The converged energies of the H$_2^+$, D$_2^+$, and HD$^+$ molecular
ions from the present Hylleraas calculations for the rovibronic
levels $(\upsilon,L)$ with $\upsilon\leq 3$ and $L\leq 3$ are listed
in Table~\ref{energy} and compared to the calculations of
Korobov~\cite{korobov06a} for H$_2^+$ and HD$^+$, who used a
different form of basis sets with pseudorandom complex exponents and
the 2002 CODATA values of the proton and deuteron
masses~\cite{mohr05a}. For the $(0,0)$ state of H$_2^+$ the present
result contains 20 significant figures, which  improves  by six
orders of magnitude the result of Korobov. Other results in
Table~\ref{energy} are converged to at least 10 significant digits.
For states $(\upsilon \geq 1, L)$ the energies are less accurate
than the corresponding $(0,L)$ states since our calculations in this
paper are for applications to ``sum over states'' determinations of
polarizabilities. Thus, the energies in Table~\ref{energy} for a
given system and value of $(\upsilon,L)$ correspond to optimized
nonlinear variational parameters for the corresponding $\upsilon=0$
state. In contrast, calculations by Korobov~\cite{korobov06a}
optimized the bases for each value $(\upsilon,L)$, and as expected,
our present values are systematically  more positive compared to
his. Recently, even more accurate energy values for $\textrm{HD}^+$
were published in Ref.~\cite{TiaTanZho12} using basis sets similar
to the present approach, but with specific optimization and
diagonalization for each separate energy level $(\upsilon,L)$.
(Accurate treatments of relativistic corrections to the ground and
first excited states were presented recently for
$\textrm{H}_2^+$~\cite{Kor08,ZhoYanShi09} and for
$\textrm{HD}^+$~\cite{Kor08,ZhoZhaYan12}.)

\subsection{Ground-state static polarizabilities and hyperpolarizabilities}\label{groundpol}

\begingroup
\squeezetable
\begin{table}[th]
\caption{Convergence of static multipole
polarizabilities $\alpha_1(0)$, $\alpha_2(0)$, and hyperpolarizability $\gamma_0(0;0,0,0)$ (in a.u.) for
the rovibronic ground-state $(\upsilon=0,L=0)$ of the H$_2^+$ ion. $N_S$, $N_P$, and $N_D$,
respectively, represent the number of
basis sets for the initial-state of $S$ symmetry, intermediate states of $P$ symmetry, and intermediate
states of $D$ symmetry.
The extrapolated values for each quantity are listed on the last line
with the computational uncertainties of the last digits in parentheses.} \label{tab1}
\begin{ruledtabular}
\begin{tabular}{llllll}
\multicolumn{2}{c}{$\alpha_1(0)$}  & \multicolumn{2}{c}{$\alpha_2(0)$} & \multicolumn{2}{c}{$\gamma_0(0;0,0,0)$}   \\
\multicolumn{1}{l}{($N_S$,$N_P$)} & \multicolumn{1}{c}{value} &\multicolumn{1}{l}{($N_S$,$N_D$)} & \multicolumn{1}{c}{value} &\multicolumn{1}{l}{($N_S$,$N_P$,$N_D$)}& \multicolumn{1}{c}{value}   \\
\hline
(420,532)      & 3.168 723 735 424 03  &(420,561)    &  1371.890 552 022 99   &(420,532,561)    &  11479.750 406 991   \\
(680,695)      & 3.168 725 614 348 09  &(680,727)    &  1371.894 443 542 72   &(680,695,727)    &  11479.793 416 663   \\
(1036,1120)    & 3.168 725 797 655 76  &(1036,954)   &  1371.894 963 825 42   &(1036,1120,954)  &  11479.795 141 858    \\
(1255,1388)    & 3.168 725 804 884 54  &(1255,1225)  &  1371.895 138 590 14   &(1255,1388,1225) &  11479.804 857 235   \\
(1504,1697)    & 3.168 725 805 220 47  &(1504,1544)  &  1371.895 140 761 38   &(1504,1697,1544) &  11479.805 065 320  \\
(1785,2050)    & 3.168 725 805 275 76  &(1785,1915)  &  1371.895 141 217 43   &(1785,2050,1915) &  11479.805 067 728   \\
(2100,2450)    & 3.168 725 805 286 34  &(2100,2342)  &  1371.895 141 236 83   &(2100,2450,2342) &  11479.805 069 686   \\
(2451,2900)    & 3.168 725 805 288 58  &(2451,2829)  &  1371.895 141 237 55   &(2451,2900,2829) &  11479.805 069 814    \\
Extrapolated   & 3.168 725 805 289(1)  &Extrapolated &  1371.895 141 24(1)    &Extrapolated     &  11479.805 07(1)   \\
\end{tabular}
\end{ruledtabular}
\end{table}
\endgroup

Table~\ref{tab1} presents a convergence study of the static
multipole polarizabilities $\alpha_1(0)$ and $\alpha_2(0)$, and the
static hyperpolarizability $\gamma_0(0;0,0,0)$ for H$_2^+$ in the
rovibronic ground-state $(\upsilon=0,L=0)$. The number of basis sets
for the state of interest is indicated by $N_S$, the number used for
the intermediate states with $P$ symmetry and $D$ symmetry are indicated
by $N_P$ and $N_D$ respectively. The extrapolated values are
obtained by assuming that the ratio between two successive
differences  stays constant as the number of basis sets used becomes
infinitely large. The static polarizabilities $\alpha_1(0)$ and
$\alpha_2(0)$ converged quickly to, respectively,  twelve and eleven
digits as the dimensions of the basis sets $N_S$, $N_P$, and $N_D$
were increased. The static hyperpolarizability, which is larger than
$\alpha_1(0)$ by four orders of magnitude, converged to the ninth
significant digit. Similar convergence tests for $\alpha_3(0)$ and
$\alpha_4(0)$ of H$_2^+$ yield the extrapolated results listed in
Table~\ref{tab2}.

\begingroup
\squeezetable
\begin{table}[th]
\caption{Static polarizabilities and
hyperpolarizabilities (in a.u.) of H$_2^+$, HD$^+$, and D$_2^+$ ions in the ground-state
$(\upsilon=0,L=0)$. The numbers in
parentheses represent the computational uncertainties obtained by extrapolation.
The first line gives the present values calculated using the CODATA 2006 masses.
The second line gives the present values calculated using $m_p=1836.152701$ and $m_d=3670.483014$
for comparison with Refs.~\cite{yan03b,hilico01a,OliPilBay12,korobov01b,bhatia00a,moss02a}.
The numbers in the square brackets for $\gamma_0$ denote powers of ten. }\label{tab2}
\begin{ruledtabular}
\begin{tabular}{llllll}
\multicolumn{1}{l}{ } & \multicolumn{3}{c}{$\alpha_1(0)$}\\\cline{2-4}
 \multicolumn{1}{l}{Author and Reference} &
\multicolumn{1}{c}{H$_2^+$} &
\multicolumn{1}{c}{D$_2^+$} &
\multicolumn{1}{c}{HD$^+$} & \\
\hline
Present\footnote{Using CODATA 2006 masses.}
    & 3.168 725 805 289(1)      & 3.071 988 697 188(1)     & 395.306 325 6742(2)\\
Present\footnote{Using  $m_p=1836.152701$, $m_d=3670.483014$}
    & 3.168 725 802 676(1)      & 3.071 988 695 66(7)      & 395.306 328 7970(6)\\
Yan \textit{et al.}~\cite{yan03b}
    &  3.168 725 802 67(1)       & 3.071 988 695 7(1)        & 395.306 328 7972(1)\\
Moss and Valenzano~\cite{moss02a}
    &                                           &                                       & 395.306 \\
Bhatia and Drachman~\cite{bhatia00a}\footnote{Using the excitation energy of the first transition from Ref.~\cite{Mos99b}.}
    &                                            &                                       & 395.289 \\
Hilico {\em et al.}~\cite{hilico01a}
   &    3.168 725 803(1)\footnote{This value, without error bar, was also obtained by Olivares Pil\'on and Baye~\cite{OliPilBay12}}             & 3.071 988 696(1)\\
Korobov~\cite{korobov01b}
   &   3.168 725 76                    & 3.071 988 68\\
Korobov~\cite{korobov01b}\footnote{Including relativistic corrections of ${\cal{O}}(\alpha^2)$}
   &   3.168 573 62                    & 3.071 838 77\\
Jacobson \textit {et al.}~\cite{jacobson00a}\footnote{Experiment}
   &   3.167 96(15)                    & 3.071 87(54) \\
\multicolumn{4}{c}{ }&\multicolumn{1}{c}{ }\\
\hline
\multicolumn{1}{l}{System (Present result)} &
\multicolumn{1}{c}{$\alpha_2(0)$} &
\multicolumn{1}{c}{$\alpha_3(0)$} &
\multicolumn{1}{c}{$\alpha_4(0)$} &
\multicolumn{1}{c}{$\gamma_0(0;0,0,0)$}  \\
\hline
H$_2^+$    & 1 371.895 141 24(1)   & 23.975 062 60(4)  & 571.963 841(2)  & 1.147 980 507(1)[4] \\
D$_2^+$    & 2 587.094 024 00(1)   & 22.890 669 73(1)  & 819.239 589(4)  & 1.967 663 142(3)[4] \\
HD$^+$      & 2 050.233 354 19(1)   & 773.42 727 01(1)  & 1434.30 534(1)  & $-$3.356 560 39(2)[9]\\
\end{tabular}
\end{ruledtabular}
\end{table}
\endgroup

The static multiple polarizabilities and hyperpolarizabilities for the
ground-state $(\upsilon=0,L=0)$ of H$_2^+$,  HD$^+$, and D$_2^+$ are  listed in Table~\ref{tab2}.
The polarizabilities and hyperpolarizabilities for the homonuclear molecular ions H$_2^+$ and D$_2^+$ have the same magnitudes.
For the heteronuclear ion HD$^+$ the corresponding
values are much larger than those for H$_2^+$ and for D$_2^+$, due to
the much smaller value of the first allowed transition energy.
Note that the hyperpolarizability of  HD$^+$  has opposite sign from H$_2^+$ and D$_2^+$ due
to the sign of the contribution from the two terms of Eq.~(\ref{hyper}).

Table~\ref{tab2} also gives a comparison with selected previous
works for the static dipole polarizabilities in the rovibronic
ground-state $(0,0)$ calculated using nonadiabatic methods (some
earlier results for $\textrm{H}_2^+$ can be found in
Ref.~\cite{taylor99a}). In order to facilitate comparison of the
present dipole polarizabilities with those of Yan \textit{et
al.}~\cite{yan03b}, we repeated the calculations by using the same
nuclear masses as they used, and the resulting values are listed in
the second line. The agreement for $\alpha_1(0)$ could hardly have
been better. However, the present static dipole polarizability of
H$_2^+$  is accurate to three parts in $10^{13}$, which improves by
one order of magnitude the result of Yan \textit{et al.} For the
static dipole polarizability of H$_2^+$, our polarizability of 3.168
725 805 289(1) is 0.025\% different from the experimental value of
3.167 96(15)~\cite{jacobson00a}. For D$_2^+$, our value is in good
agreement with the less accurate result of Hilico \textit{et
al.}~\cite{hilico01a} and slightly larger than the result of Yan
\textit{et al.}~\cite{yan03b}. The present dipole polarizability
3.071 988 697 188(1) of D$_2^+$ agrees with the experimental value
3.07187(54) at the level of 0.004\%. For HD$^+$, our result is much
more accurate than the early result of Moss and
Valenzano~\cite{moss02a}. Some other nonadiabatic calculations of
the quadrupole (and higher order) polarizabilities are given in
Refs.~\cite{yan03b,OliPilBay12} and we are in good agreement. There
is a previous nonadiabatic calculation of the second
hyperpolarizability for H$_2^+$: Moss and Valenzano~\cite{moss02a}
find $\gamma=1.14\times 10^4$, in harmony with our result.

It is interesting to examine in more detail the quadrupole
polarizibility and second hyperpolarizability calculations with
previous Born-Oppenheimer treatments, where the quantities are
separated into ``electronic", ``vibrational", and ``rotational"
contributions~\cite{BisLam88}. As exhibited in Table~\ref{tab2}, the
relative magnitude of $\alpha_2(0)$ is much larger than those of
$\alpha_1(0)$ and $\alpha_3(0)$, which is related to the available
low-lying virtual state in the energy denominator (a similar
argument pertains to $\alpha_4(0)$). In the Born-Oppenheimer
approach, the virtual excitation corresponds to no change in the
electronic or vibrational quantum number, but a change in the
rotational quantum number by 2. Bishop and Lam~\cite{BisLam88}, (see
their table~7), found $\alpha_2(0)=1370.7$ a.u., composed of
electronic, vibrational, and rotational contributions of,
respectively, $4.8$ a.u., $3.69$ a.u., and $1362.24$ a.u., where the
relatively larger rotational contribution reflects the low-lying
virtual excitation. In a recent paper, Olivares Pil\'{o}n and
Baye~\cite{OliPilBay12} compared their total nonadiabatic
calculation of $\alpha_2(0)$ for the ground state to a second order
perturbation theoretic sum over the first four vibrational states
(their Eq.~(25)) using matrix elements from their nonadiabatic
calculation. They found that the nonadiabatic result was  greater by
an additive factor of $4.793$, compared to the summation and
attributed this to ``the contribution of the continuum." In the
language of Bishop and Lam~\cite{BisLam88}, the summation
corresponds to including most of the ``vibrational" and
``rotational" components of $\alpha_2(0)$. The missing quantity is
supplied by Bishop and Lam's  ``electronic" component of $4.8$.
Evidently, the partial sum of Olivares Pil\'{o}n and Baye does not
converge to the correct value simply because of the neglect of
higher electronic excitations.

The magnitude of the static hyperpolarizability can also be understood along similar lines in Born-Oppenheimer picture.
Earlier work using finite field methods by Bishop and Solunac~\cite{BisSol85} and by Adamowicz and Bartlett~\cite{AdaBar86} established
that nonadiabatic effects were not the source of the large hyperpolarizability.
Subsequently, Bishop and Lam~\cite{BisLam88} calculated $\gamma(0)=11537.16$, with
electronic, vibrational, and rotational contributions of, respectively,
$29.76$, $568.7$, and $10945.13$, where again the larger rotational contribution
is mainly due to the virtual transition where the rotational quantum number changes by 2.

Dynamic hyperpolarizabilities  pertain to the four nonlinear optical processes~(cf. Refs.~\cite{OrrWar71,pipin92a,SheRic94}):
Thus, the quantity $\gamma_0(-\omega;\omega,0,0)$ is the dc Kerr effect, $\gamma_0(-\omega;\omega,\omega,-\omega)$ represents  degenerate four-wave mixing (DFWM), $\gamma_0(-2\omega;0,\omega,\omega)$ is  electric-field-induced second-harmonic generation (ESHG) and $\gamma_0(-3\omega;\omega,\omega,\omega)$ is  third-harmonic generation (THG).

In the Born-Oppenheimer approach, the rotational contributions to the dynamic hyperpolarizabilities
for the dc Kerr and DFWM processes at optical wavelengths are expected to be comparable to  $\gamma(0)$
while the rotational contributions to the ESHG and THG processes are expected to be much reduced in comparison to $\gamma(0)$~\cite{SheRic94}.
For $\textrm{H}_2^+$, we calculated the dc Kerr, DFWM, and ESHG hyperpolarizabilities at a wavelength of 632.8~nm.
Using the available Born-Oppenheimer
calculations of the electronic contributions from Bishop and Lam~\cite{BisLam87} (their tables 2--4) (at the
$\textrm{H}_2^+$ equilibrium internuclear distance 2 a.u.) and the vibrational contributions
(their table 7), we  estimated the rotational contributions by subtraction from our nonadiabatic values.
The results are given in Table~\ref{compare-hyper}.
The nonadiabatic calculations were carried out using the methods described herein with
the largest basis set $(N_s, N_p, N_d) = (2840, 2900, 2829)$ and were converged values.
(Unfortunately, we were unable to obtain a converged value for THG at this wavelength.)
Nevertheless, the results yield estimates of the rotational components
of dc Kerr and DFWM that are comparable to the static value.
For example, at 632.8~nm (He-Ne laser), we find that the dc Kerr rotational contribution is around $3787$
compared to the ESHF rotational contribution of $-35$.
\begin{table}[th]
\caption{For $\mbox{H}_2^+$, estimation of rotational contributions to the dc Kerr, DFWM, and ESHG processes,
in the Born-Oppenheimer picture, using tabulated electronic and vibrational values and
the present nonadiabatic values, at wavelength of 632.8~nm.
The values for the ``Electronic'' component'' correspond to the internuclear distance of 2 a.u.
}\label{compare-hyper}
\begin{ruledtabular}
\begin{tabular}{llll}
\multicolumn{1}{l}{component} & \multicolumn{1}{c}{dc Kerr} & \multicolumn{1}{c}{DFWM}
& \multicolumn{1}{c}{ESHG}  \\
\hline
Nonadiabatic (total)                               & 4028.6 & 8445.1 & 14.631 \\
Electronic (Ref.~\protect\cite{BisLam87}) &  54.3     & 56.2    &   58.3     \\
Vibrational (Ref.~\protect\cite{BisLam87})           & 187.21 & 388.87 & -8.65  \\
\hline
Rotational (row 1-(row 2+row3))            &  3787   & 8000   &  -35.0    \\
\end{tabular}
\end{ruledtabular}
\end{table}

\subsection{Dynamic dipole polarizabilities and hyperpolarizabilities for the rovibronic ground-state of HD$^+$ }

\begingroup
\squeezetable
\begin{table}[th]
\caption{Dynamic dipole polarizabilities and hyperpolarizabilities (in a.u.)
for HD$^+$ in the  ground-state $(\upsilon=0,L=0)$   for photon
energies $\omega\leq 0.0004$~a.u. The numbers
in parentheses represents the computational
uncertainties. The numbers in
the square brackets denote powers of ten.}\label{tab3}
\begin{ruledtabular}
\begin{tabular}{llllll}
\multicolumn{1}{c}{$\omega \times 10^{4}$}
 & \multicolumn{1}{c}{$\alpha_1(\omega)$} & \multicolumn{1}{c}{dc Kerr} & \multicolumn{1}{c}{DFWM} & \multicolumn{1}{c}{ESHG} & \multicolumn{1}{c}{THG}\\
 \hline
0.2    & 399.273 277 88(1)       & $-$3.41610834(1)[9]    &$-$3.47659308(2)[9]     &$-$3.54049773(2)[9]     & $-$3.74430183(2)[9] \\
0.4    & 411.670 828 83(1)       & $-$3.60545269(2)[9]    &$-$3.87061373(2)[9]     &$-$4.20449102(2)[9]     & $-$5.57473833(2)[9]\\
0.6    & 434.153 094 36(1)       & $-$3.96135711(2)[9]    &$-$4.66118020(2)[9]     &$-$5.89905124(3)[9]     & $-$2.04245391(1)[10]\\
0.8    & 470.133 273 96(1)       & $-$4.56456592(2)[9]    &$-$6.14508585(2)[9]     &$-$1.159595160(4)[10]   & 9.78040986(3)[9]\\
1.0    & 526.283 388 35(1)       & $-$5.58859503(3)[9]    &$-$9.05764398(4)[9]     &2.99296605(1)[12]       & 4.05178750(1)[9]  \\
1.2    & 616.411 116 77(1)       & $-$7.44289693(3)[9]    &$-$1.549390983(5)[10]   &1.310381956(4)[10]      & 2.816792369(6)[9]\\
1.4    & 773.207 487 30(1)       & $-$1.128714230(5)[10]  &$-$3.304470220(9)[10]   &7.89267238(2)[9]        & 2.506628276(6)[9]  \\
1.6    & 1095.46 735 839(1)      & $-$2.165280366(8)[10]  &$-$1.033521867(2)[11]   &7.29733829(2)[9]        & 2.799248441(7)[9] \\
1.8    & 2081.02 392 881(1)      & $-$7.39271980(2)[10]   &$-$7.964170683(8)[11]   &1.042204934(1)[10]      & 4.53270763(2)[9] \\
2.0    & $-$245402.484 3(1)      & $-$9.60586446(3)[14]   &1.51083701452(7)[18]    &$-$1.016296893(1)[12]   & $-$4.88278453(2)[11]    \\
2.2    & $-$1846.878 075 95(3)   & $-$5.00377102(2)[10]   &7.760677632(7)[11]      &$-$6.877974834(7)[9]    & $-$3.60827031(3)[9]    \\
2.4    & $-$883.279 541 49(2)    & $-$1.030699336(3)[10]  &1.090946710(2)[11]      &$-$3.253912120(4)[9]    & $-$1.85432993(3)[9]    \\
2.6    & $-$562.828 160 15(2)    & $-$3.659961421(7)[9]   &4.08371648(1)[10]       &$-$2.367173671(6)[9]    & $-$1.46202643(3)[9] \\
2.8    & $-$403.893 734 87(1)    & $-$1.577815323(2)[9]   &2.94181712(1)[10]       &$-$2.67134745(1)[9]     & $-$1.78534388(5)[9]   \\
3.0    & $-$309.565 802 26(1)    & $-$7.219101053(4)[8]   &$-$2.874893746(7)[11]   &3.66561097(3)[10]       & 2.6441199(1)[10]\\
3.2    & $-$247.474 914(1)       & $-$3.120300419(3)[8]   &$-$5.555601770(5)[9]    &9.3765350(2)[8]         & 7.2597679(4)[8]\\
3.4    & $-$203.741 252(1)       & $-$9.55821852(5)[7]    &$-$1.473800328(4)[9]    &3.20406687(8)[8]        & 2.6341471(2)[8]\\
3.6    & $-$171.427 953(1)       & 2.69257734(6)[7]       &$-$5.41613704(5)[8]     &1.51419712(6)[8]        & 1.2971185(1)[8]\\
3.8    & $-$146.685 806(1)       & 1.00270127(1)[8]       &$-$2.30385033(5)[8]     &8.4768476(4)[7]         & 7.3438222(8)[7]\\
4.0    & $-$127.209 551(1)       & 1.46714081(1)[8]       &$-$1.05875887(4)[8]     &5.3872799(4)[7]         & 4.5225693(6)[7] \\
\end{tabular}
\end{ruledtabular}
\end{table}
\endgroup

Since the transition $(0,0)\rightarrow (0,1)$ is a forbidden transition for the H$_2^+$ and D$_2^+$ ions,
the first allowed transitions are at about $\omega=0.1903$~a.u. for H$_2^+$ and $\omega=0.2004$~a.u. for D$_2^+$,
corresponding to  ``electronic transitions'' (in the Born-Oppenheimer picture) and
which are not in the visible spectrum.
Thus, in this subsection we concentrate only on the dynamic dipole polarizability and hyperpolarizability of the
HD$^+$ system, for which optical transitions can occur.
Table~\ref{tab3} presents selectively some values of dynamic dipole polarizabilities and hyperpolarizabilities for ground-state
 HD$^+$. All of the values are accurate to at least nine significant figures.
The effect of  the $(0,0)$ to $(0,1)$ resonance near the energy $2.0
\times 10^{-4}$, see Table~\ref{energy}, on the quantities tabulated
is apparent.

\begin{figure}
\includegraphics[width=0.7\textwidth]{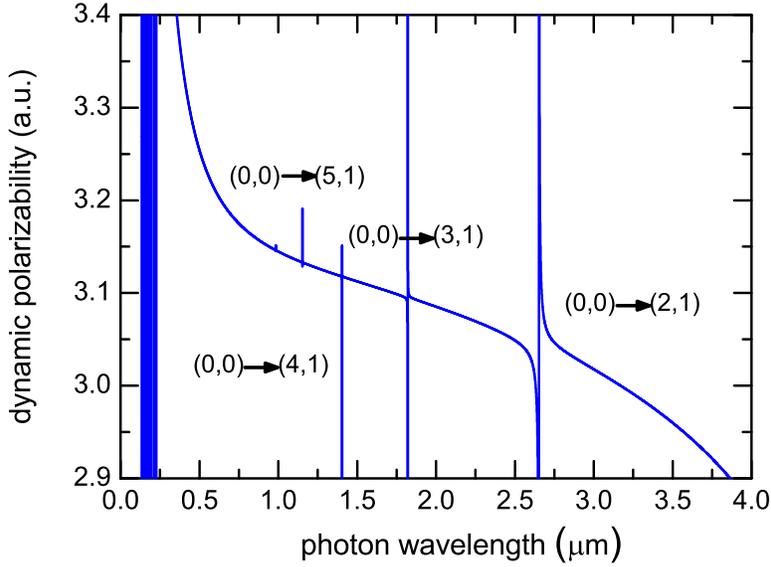}
\caption{(Color online) Dynamic dipole polarizability
$\alpha_1(\omega)$ (in a.u.) for the rovibronic ground-state $(\upsilon=0,L=0)$ of the HD$^+$ ion for
photon wavelengths  from 0  to 4 $\mu\textrm{m}$. The resonances $(0,0)\rightarrow (\upsilon,1)$  in the dynamic polarizability
are marked. } \label{f3}
\end{figure}

\begin{figure}
\includegraphics[width=0.7\textwidth]{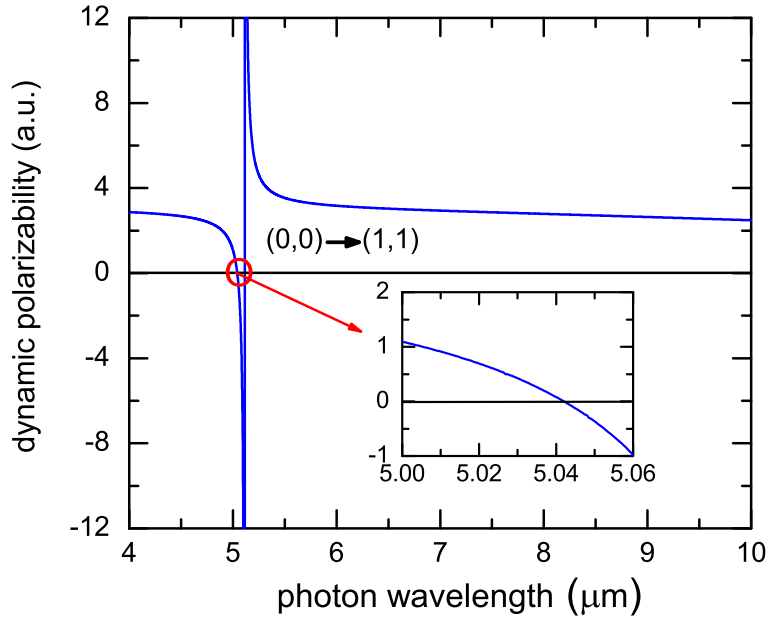}
\caption{(Color online) Dynamic dipole polarizability
$\alpha_1(\omega)$ (in a.u.) for the rovibronic ground-state $(\upsilon=0,L=0)$ of  the HD$^+$ ion for
photon wavelengths from 4 to 10 $\mu\textrm{m}$. The resonance $(0,0)\rightarrow (1,1)$ in the dynamic polarizability
 is marked. In the inset the region where $\alpha_1(\omega)=0$ around $5.04$~$\mu\textrm{m}$
 is shown in greater detail.}\label{f4}
\end{figure}

\begin{figure}
\includegraphics[width=0.7\textwidth]{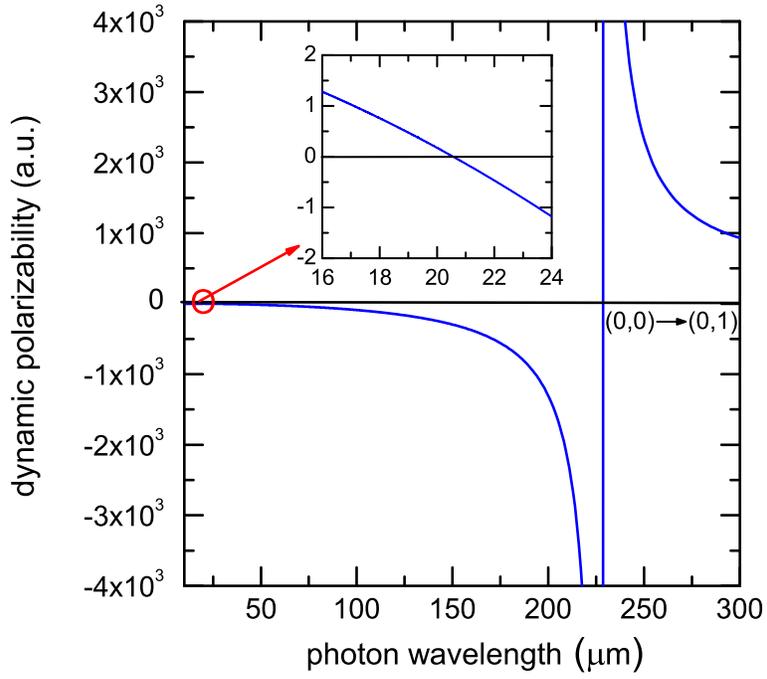}
\caption{(Color online) Dynamic dipole polarizability
$\alpha_1(\omega)$ (in a.u.) for the rovibronic ground-state $(\upsilon=0,L=0)$ of the
HD$^+$ ion for photon wavelengths from 10 to 300 $\mu\textrm{m}$.
The $(0,0)\rightarrow (0,1)$ resonance in the dynamic polarizability is marked.
The inset is a magnification of the circled position around 20~$\mu\textrm{m}$
where $\alpha_1(\omega)=0$.}\label{f5}
\end{figure}

\begin{figure}
\includegraphics[width=0.7\textwidth]{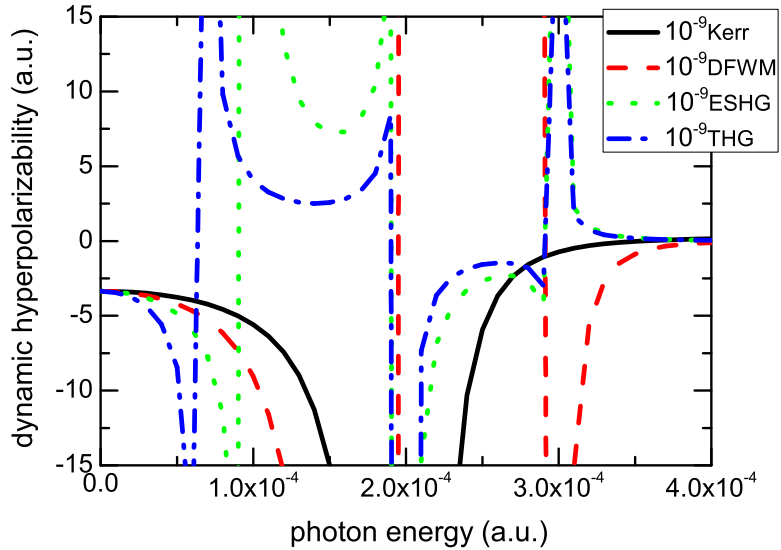}
\caption{(Color online) Dynamic hyperpolarizabilities (in a.u.), see
text, for the rovibronic ground-state $(\upsilon=0,L=0)$ of the
HD$^+$ ion for photon energies $\omega \leq 0.0004$~a.u. The solid
black line represents the Kerr effect, the dashed red line denotes
DFWM, the dotted green line represents ESHG, and the dash-dot blue
line is THG.}\label{f6}
\end{figure}

Figs.~\ref{f3}--\ref{f5} show the dynamic dipole polarizability
$\alpha_1(\omega)$ of HD$^+$ in the  ground state as a function of
wavelength $\lambda =c/\omega$ in $\mu\textrm{m}$. The perpendicular
lines represent the positions of resonant transitions. That there
are many resonance transitions as $\lambda\rightarrow
0$~$\mu\textrm{m}$ is evident in  Fig~\ref{f3}. However, for the
wavelengths $\lambda=4-10$~$\mu\textrm{m}$, shown in Fig.~\ref{f4},
and the wavelengths $\lambda=10-300$~$\mu\textrm{m}$, shown in
Fig.~\ref{f5}, there is only one transition in each range. In the
inserts for Figs.~\ref{f4} and \ref{f5} the plots are magnified to
show the positions where $\alpha_1(\omega)=0$. In Fig.~\ref{f4}, the
transition $(0,0)\rightarrow (1,1)$ occurs at
$\lambda=5.115454421$~$\mu\textrm{m}$ (or photon energy of 0.008907
a.u.) and $\alpha_1(\omega)=0$  at
$\lambda=5.05024967$~$\mu\textrm{m}$ (0.009022 a.u.). In
Fig.~\ref{f5}, the transition $(0,0)\rightarrow (0,1)$ occurs  at
$\lambda=227.816763$~$\mu\textrm{m}$ and $\alpha_1(\omega)=0$ occurs
at $\lambda=20.5147918$~$\mu\textrm{m}$. Our results for the $(0,0)$
state are in good agreement with the less accurate results of
Koelemeij~\cite{koelemeij11a}, who combined the nonadiabatic
polarizability calculations of Moss and Valenzano~\cite{moss02a}
with vibrational-rotational energies and electric dipole matrix
elements  calculated in the Born-Oppenheimer picture to obtain
values of $\alpha_1(\omega)$ in the infrared. In Fig.~\ref{f6} the
various hyperpolarizabilities (dc Kerr, DFWM, ESHG, and THG) are
plotted over the energy range $0<\omega<4\times 10^{-4}$ a.u. The
first resonant transition is prominent near $2.0\times 10^{-4}$ a.u.
Note that sign changes for ESHF and THG occur at lower energies and
sign changes for DFWM, ESHG, and THG occur at higher energies as
well, due to the complicated perturbation theoretic expressions.

\subsection{First excited-state static polarizabilities and hyperpolarizabilities}\label{excitedpol}

\begingroup
\squeezetable
\begin{table}[th]
\caption{Convergence of static scalar and tensor dipole
polarizabilities (in a.u.) for the first excited-state
$(\upsilon=0,L=1)$ of the H$_2^+$ ion.} \label{tab4}
\begin{ruledtabular}
\begin{tabular}{lccccc}
\multicolumn{1}{c}{($N_S$,$N_P$,$N_{(pp')P}$,$N_D$)} & \multicolumn{1}{c}{$\alpha_1(S)$} & \multicolumn{1}{c}{$\alpha_1((pp')P)$} & \multicolumn{1}{c}{$\alpha_1(D)$} & \multicolumn{1}{c}{$\alpha_1(0)$} & \multicolumn{1}{c}{$\alpha_1^{(T)}$(0)} \\
\hline
(124,140,185,131)      & 0.650 846 694 354 224  & 0.599 484 436 191    & 1.903 824 885 459  & 3.154 156 016 01  & $-$0.541 486 964 805\\
(175,205,255,295)      & 0.650 848 412 543 154  & 0.610 162 196 819    & 1.913 244 494 929  & 3.174 255 104 29  & $-$0.537 091 763 627\\
(240,290,342,404)      & 0.650 848 769 703 336  & 0.612 800 029 560    & 1.913 969 508 078  & 3.177 618 307 34  & $-$0.535 845 705 731\\
(420,532,448,561)      & 0.650 848 734 133 560  & 0.613 242 706 580    & 1.914 080 822 144  & 3.178 172 262 86  & $-$0.535 635 463 058\\
(680,695,575,727)      & 0.650 848 734 170 854  & 0.613 340 747 207    & 1.914 098 088 958  & 3.178 287 570 34  & $-$0.535 588 169 463\\
(1036,1120,725,954)    & 0.650 848 734 188 369  & 0.613 352 442 310    & 1.914 098 306 929  & 3.178 299 483 43  & $-$0.535 582 343 726\\
(1255,1388,900,1225)   & 0.650 848 734 193 030  & 0.613 353 636 014    & 1.914 100 742 509  & 3.178 303 114 90  & $-$0.535 581 990 437\\
(1504,1697,1102,1544)  & 0.650 848 734 193 204  & 0.613 353 828 247    & 1.914 100 852 289  & 3.178 303 414 73  & $-$0.535 581 905 299\\
(1785,2050,1333,1915)  & 0.650 848 734 193 242  & 0.613 353 844 426    & 1.914 100 898 193  & 3.178 303 476 81  & $-$0.535 581 901 799\\
(2100,2450,1595,2342)  & 0.650 848 734 193 251  & 0.613 353 845 680    & 1.914 100 900 765  & 3.178 303 480 64  & $-$0.535 581 901 430\\
(2451,2900,1890,2829)  & 0.650 848 734 193 253  & 0.613 353 845 788    & 1.914 100 901 118  & 3.178 303 481 10  & $-$0.535 581 901 411\\
Extrapolated           &                      &                      &                    & 3.178 303 481(1)  & $-$0.535 581 901 4(1)\\
\end{tabular}
\end{ruledtabular}
\end{table}
\endgroup

Table~\ref{tab4} shows a convergence study of the static scalar and
tensor dipole polarizabilities for H$_2^+$ in the rovibronic
excited-state $(\upsilon=0,L=1)$. The integer $N_{(pp')P}$
represents the number of intermediate states used when the electron
and one nucleus are both in excited states of $p$ symmetry to form
the total angular momentum $L=1$. The contribution of the
configuration $\alpha_1((pp')P)$ to $\alpha_1(0)$ is about 20\%, as
shown in Table~\ref{tab4}. The final static scalar and tensor dipole
polarizabilities are both converged to the ninth figures.
Calculations of $\alpha_1(0)$ for $\textrm{D}_2^+$ were also carried
out  with similar results. Results for the static scalar and tensor
dipole polarizabilities for HD$^+$ are presented in Table~\ref{HD}
and there is a partial cancellation between two intermediate
symmetries, which can be seen by comparing columns~2 and 4. For the
largest basis set, $\alpha_1(S)=-130.024\,382\,526\,724$~a.u and
$\alpha_1(D)=133.405\,246\,966\,154$~a.u.; thus, when the two terms
are added a loss of two significant figures results. Similar
calculations were performed to obtain the static multipole
polarizabilities $\alpha_2(0)$ and $\alpha_3(0)$  of the H$_2^+$,
D$_2^+$, and HD$^+$ ions in their first  excited-states
$(\upsilon=0,L=1)$. Our results for  $\alpha_1(0)$ and
$\alpha_1^{(T)}(0)$ for all three molecular ions are in agreement
with the recent results of Schiller \textit{et
al}~\cite{SchBakKor14b}, which are accurate to 8 significant digits.
%
\begingroup
\squeezetable
\begin{table}[th]
\caption{Convergence of static scalar and tensor dipole polarizabilities (in a.u.) for HD$^+$ in the rovibronic excite-state $(\upsilon=0,L=1)$.} \label{HD}
\begin{ruledtabular}
\begin{tabular}{llllll}
\multicolumn{1}{c}{($N_S$,$N_P$,$N_{(pp')P}$,$N_D$)} & \multicolumn{1}{c}{$\alpha_1(S)$} & \multicolumn{1}{c}{$\alpha_1((pp')P)$} & \multicolumn{1}{c}{$\alpha_1(D)$} & \multicolumn{1}{c}{$\alpha_1(0)$} & \multicolumn{1}{c}{$\alpha_1^{(T)}$(0)} \\
\hline
(124,140,104,150)       & -130.074 770 552 639   & 0.547 263 971 717    & 133.368 575 021 506   & 3.841 068 441   & 117.011 545 036 \\
(240,290,221,325)       & -130.024 339 444 310   & 0.600 501 791 517    & 133.405 220 142 988   & 3.981 382 490   & 116.984 068 326 \\
(420,532,406,616)       & -130.027 278 013 066   & 0.600 795 791 278    & 133.404 624 757 699   & 3.978 142 536   & 116.987 213 433 \\
(680,890,675,815)       & -130.024 394 969 388   & 0.608 486 409 588    & 133.405 156 819 278   & 3.989 248 259   & 116.988 122 492 \\
(1036,1388,1044,1055)   & -130.024 382 831 359   & 0.609 081 936 428    & 133.405 235 157 018   & 3.989 934 262   & 116.988 400 284 \\
(1504,1697,1271,1340)   & -130.024 382 564 424   & 0.609 175 971 976    & 133.405 245 134 960   & 3.990 038 543   & 116.988 446 037 \\
(1785,2050,1529,1674)   & -130.024 382 532 136   & 0.609 261 536 706    & 133.405 246 685 012   & 3.990 125 689   & 116.988 488 632 \\
(2100,2450,1820,2061)   & -130.024 382 527 321   & 0.609 275 874 989    & 133.405 246 931 565   & 3.990 140 279   & 116.988 495 772 \\
(2451,2900,2299,2505)   & -130.024 382 526 724   & 0.609 281 615 158    & 133.405 246 966 154   & 3.990 146 055   & 116.988 498 638 \\
Extrapolated            &                        &                      &                       & 3.990 148(2)    & 116.988 499(1)\\
\end{tabular}
\end{ruledtabular}
\end{table}
\endgroup
%
%
\begin{center}
\begin{table}
\caption{Static polarizabilities and hyperpolarizabilities (in a.u.)
of H$_2^+$, HD$^+$, and D$_2^+$ ions in the first excited-state
$(\upsilon=0,L=1)$. The numbers in parentheses represent the
computational uncertainties. The numbers in the square brackets
denote powers of ten. }\label{tab5}
\begin{tabular}{llllll}
\hline\hline
 \multicolumn{1}{l}{System} &
\multicolumn{1}{c}{$\alpha_1(0)$} &
\multicolumn{1}{c}{$\alpha_1^{(T)}(0)$} & \multicolumn{1}{c}{$\alpha_2(0)$} &
\multicolumn{1}{c}{$\alpha_3(0)$} \\
\hline
H$_2^+$      & 3.178 303 481(1)   & $-$0.535 581 901 4(1)    & 505.648 042 6(1)      & 24.076 096(1)  \\
                    & 3.178 303 479\footnote{Ref.~\cite{OliPilBay12}} &&&\\
D$_2^+$      & 3.076 590 373(1)  & $-$0.505 301 361 2(1)    & 942.776 985 8(1)      & 22.938 665(1)  \\
HD$^+$        & 3.990 148(2)         &   116.988 499(1)             & 751.719 465 6(3)      & 1265.003 2(2) \\
                    & 3.990 667\footnote{Ref.~\cite{moss02a}} &&&\\
\hline
\multicolumn{1}{l}{System} &
\multicolumn{1}{c}{$\gamma_0(0)$} &
\multicolumn{1}{c}{$\gamma_2(0)$}  \\
\hline
H$_2^+$          & 4580.48(3)          & $-$835.88(2)\\
D$_2^+$          & 7486.2986(1)      & $-$1228.7041(1)\\
HD$^+$            & 1.0708026(2)[9] & $-$1.1823956(1)[9]\\
\hline\hline
\end{tabular}
\end{table}
\end{center}

Table~\ref{tab5} summarizes the final values of the static multipole
polarizabilities and hyperpolarizabilities for the H$_2^+$, D$_2^+$ and
HD$^+$ ions in their first excited-states $(\upsilon=0,L=1)$.
From this table, we can see that  dipolar and octupolar quantities for HD$^+$ are much larger
than those for H$_2^+$ and D$_2^+$, especially for the
hyperpolarizability, due to the allowed low-lying virtual state
entering in the $\mbox{HD}^+$ case.
For $\mbox{HD}^+ (0,1)$, Moss and Valenzano~\cite{moss02a} found $\alpha_1(0) = 3.990\,667$
in a nonadiabatic calculation.
For $\mbox{H}_2^+ (0,1)$,  Bishop and Lam~\cite{BisLam88}
find $\gamma_0 = 4\,634.39$ in the Born-Oppenheimer approximation.

\subsection{Static Stark shift}

The static Stark shift $\Delta E$ for the rovibronic ground-state $(0,0)$ of a hydrogen molecular ion in
an electric field of strength ${\cal{E}}$ is
\begin{eqnarray}
\Delta E=-\frac{{\cal{E}}^2}{2}\alpha_1(0)-\frac{{\cal{E}}^4}{24}\gamma_0(0;0,0,0)
\end{eqnarray}
and the relative ratio between the second term and the first term is written as
\begin{eqnarray}
X=\frac{\gamma_0(0;0,0,0){\cal{E}}^2}{12\alpha_1(0)} .
\end{eqnarray}
This ratio determines the extent to which the Stark shift is influenced by the hyperpolarizability at high field strengths. Using the  values of Table~\ref{tab2}, at ${\cal{E}}=6.67\times 10^{-5}$~a.u. $\sim(334\; \textrm{kV/cm})$, we find $X=1.3\times 10^{-6}$ for H$_2^+$,
$X=2.4\times 10^{-6}$ for D$_2^+$, and $X=-0.0031$ for HD$^+$.
When ${\cal{E}}= 2.11\times 10^{-4}$~a.u. $\sim(1087\; \textrm{kV/cm})$, we find $X=1.3\times 10^{-5}$ for H$_2^+$, $X=2.4\times 10^{-5}$ for D$_2^+$,  and $X=-0.032$ for HD$^+$.
So the hyperpolarizability effect  is more significant
for the HD$^+$ system compared to either the H$_2^+$ or D$_2^+$ system. In particular, it can cancel the Stark shift from the dipole polarizabilities.

The leading term of static Stark shift $\Delta E$ for the transition $(0,0)\rightarrow (0,1)$ of hydrogen molecular ions in the electric field strength ${\cal{E}}$ is
\begin{eqnarray}
\Delta E =-\frac{{\cal{E}}^2}{2}\big[\alpha_1^{(0,0)}(0)-\alpha_1^{(0,1)}(0)\big],
\end{eqnarray}
where $\alpha_1^{(0,0)}(0)$ and $\alpha_1^{(0,1)}(0)$ represent the
static dipole polarizabilities for the ground-state $(0,0)$ and
excited-state (0,1) respectively. Using the present values from
Tables~\ref{tab2} and \ref{tab5}, we  obtain  $\Delta
\alpha_1(0)=\alpha_1^{(0,0)}(0)-\alpha_1^{(0,1)}(0)=-0.009\,577\,675\,711$~a.u.
for H$_2^+$, $\Delta \alpha_1(0)=-0.004\,601\,675\,812$~a.u. for
D$_2^+$, and $\Delta \alpha_1(0)=391.316\,177\,674 \,2$~a.u. for
HD$^+$. Thus the  second-order Stark shift will be larger for HD$^+$
than for either the H$_2^+$ or D$_2^+$ ion.

\subsection{Tune-out and magic wavelengths of HD$^+$ }\label{magictuneout}

At certain laser frequencies where the dynamic polarizability vanishes it may be possible to eliminate the shift induced by an applied
laser field~\cite{LeBThy07}---these frequencies are known as \textit{tune-out} frequencies
or wavelengths.
In addition, there might exist laser frequencies for an
ion in two different states where the radiation induced shifts are equal (because the dynamic polarizabilities
are equal at those frequencies): These frequencies are known as \textit{magic} frequencies or wavelengths.

\begin{figure}
\includegraphics[width=0.7\textwidth]{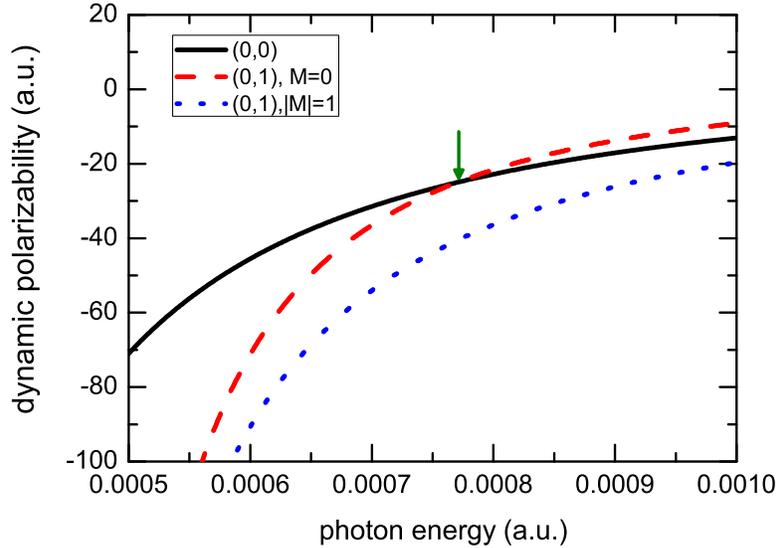}
\caption{(Color online) Dynamic dipole polarizabilities
$\alpha_1(\omega)$ (in a.u.) of HD$^+$ for photon energies between
0.0005 and 0.001 a.u. The solid black line denotes the dynamic
polarizabilities of ground state $(\upsilon=0,L=0)$. The dashed red
and dotted blue lines represent the dynamic polarizabilities of the
first excited state$(\upsilon=0,L=1)$ with $M=0$ and $|M|=1$
respectively. The magic-wavelength for the transition
$(0,0)\rightarrow (0,1)$ is marked by the arrow. } \label{magic1}
\end{figure}

\begin{figure}
\includegraphics[width=0.7\textwidth]{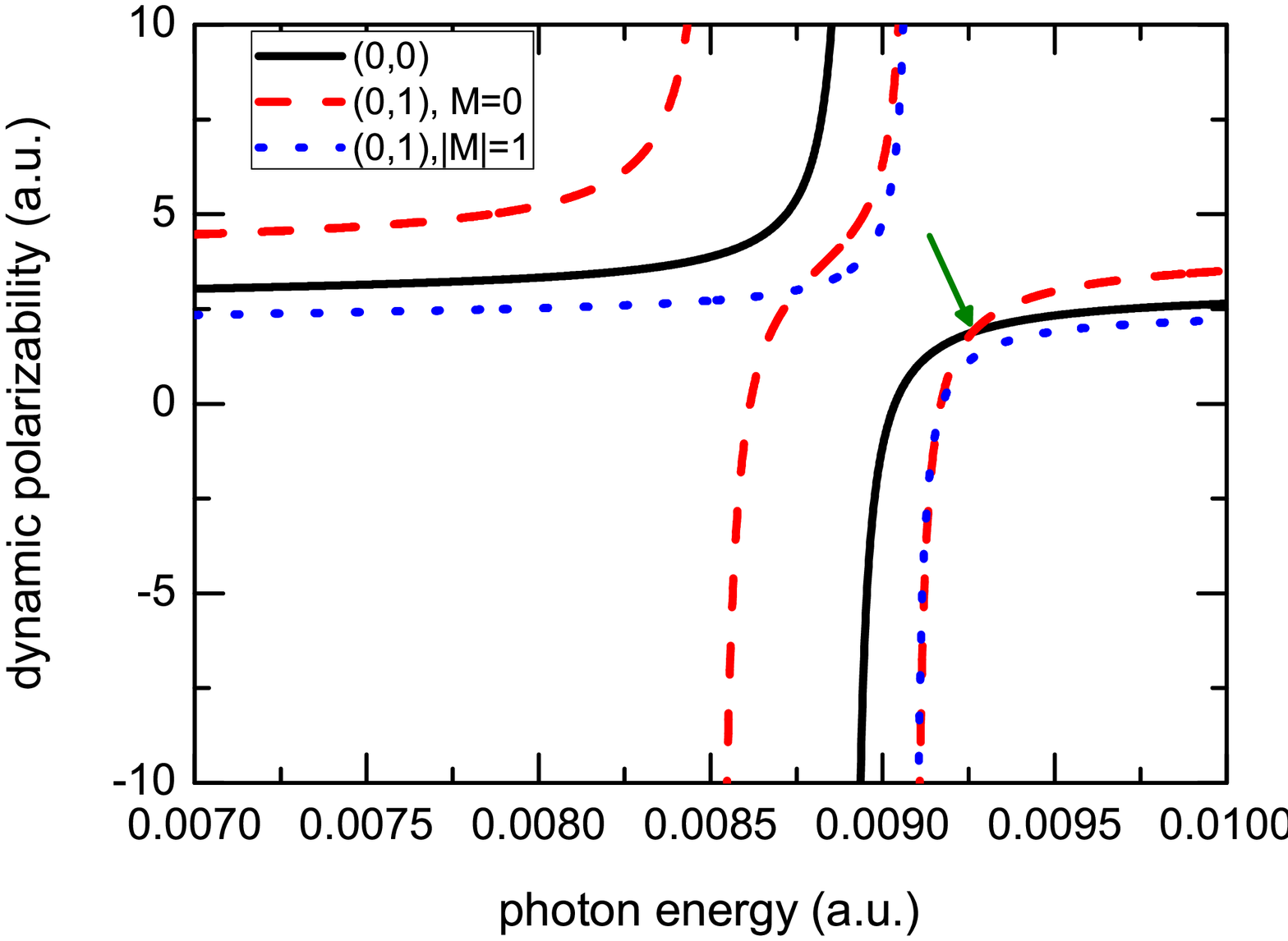}
\caption{(Color online) As for Fig.~\protect\ref{magic1}, but
for photon energies between 0.007 and 0.01 a.u.} \label{magic2}
\end{figure}

\begin{figure}
\includegraphics[width=0.7\textwidth]{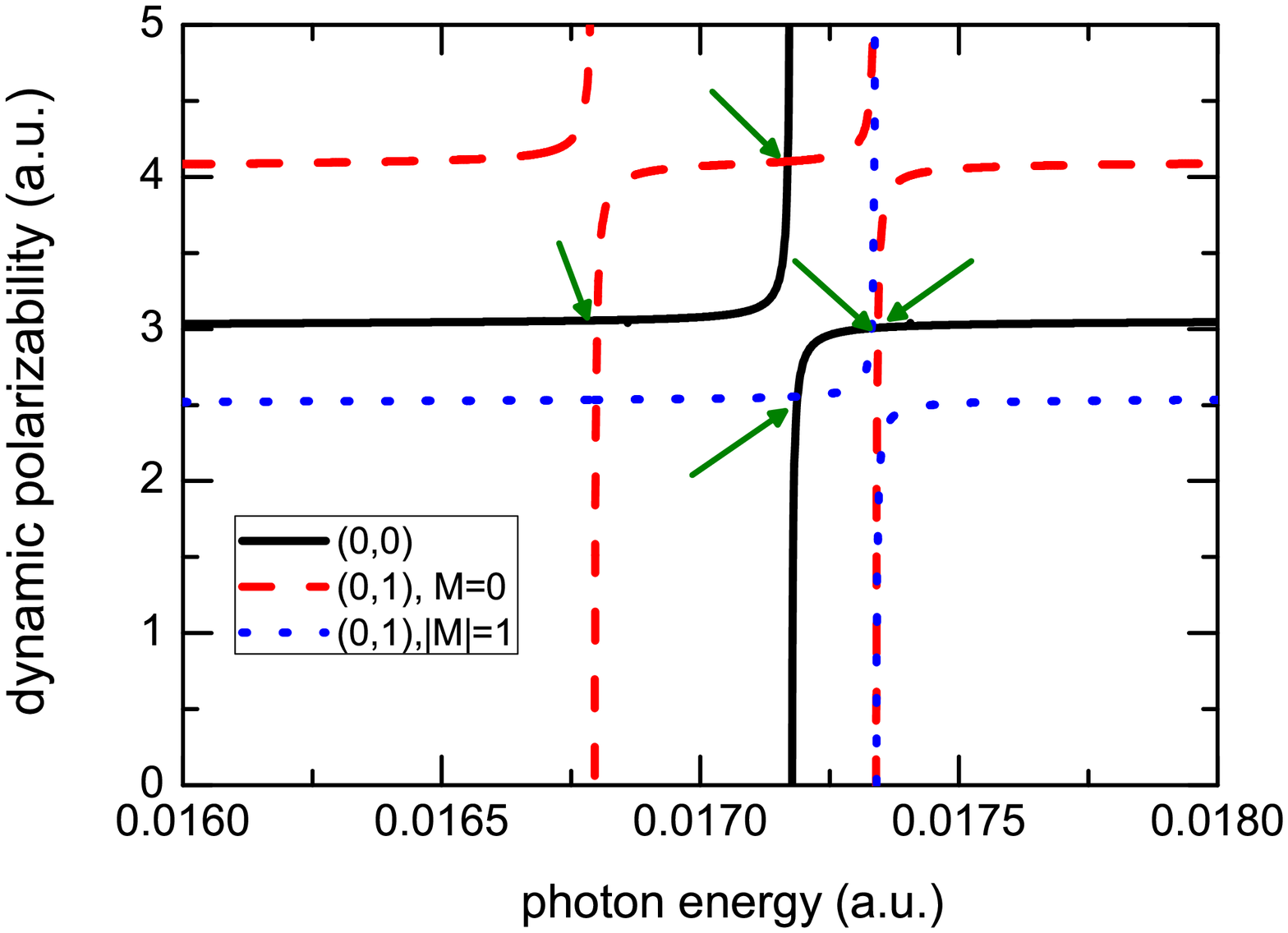}
\caption{(Color online) As for Fig.~\protect\ref{magic1}, but
for photon energies between 0.016 and 0.018 a.u. } \label{magic3}
\end{figure}

For the first excited-state $(0,1)$ of HD$^+$, the dynamic dipole polarizability is
\begin{equation}
\alpha_{1,M}(\omega)=\alpha_1(\omega)+\alpha_1^T(\omega)\frac{3M^2-L(L+1)}{L(2L-1)},
\end{equation}
where $M$ is the magnetic quantum number. In
Table~\ref{tune-out-tab} we list some of low-lying (in energy)
tune-out wavelengths for the ground state and the first excited
state of HD$^+$. The positions of magic-wavelengths between the
ground-state and the first excited-state of HD$^+$ are marked by the
arrows in Figs.~\ref{magic1}--\ref{magic3}, there are no
magic-wavelengths in the visible light range. In
Table~\ref{magic-tab} we list the values of the magic wavelengths
indicated in Figs.~\ref{magic1}--\ref{magic3}.

\begingroup
\squeezetable
\begin{table}[th]
\caption{Tune-out wavelengths (in a.u.) for HD$^+$. The numbers in parentheses represent the
computational uncertainties. }\label{tune-out-tab}
\begin{ruledtabular}
\begin{tabular}{lcllll}
\multicolumn{1}{l}{ State } &\multicolumn{1}{c}{$M$}&\multicolumn{4}{c}{Tune-out wavelengths} \\
 \hline
$(0,0)$  & 0     & 0.002 215 386 568(1) & 0.009 036 752 923(1) & 0.017 178 225 41(1)\\
$(0,1)$  & 0     & 0.001 578 607 28(1)  & 0.008 614 811 326(1) & 0.009 169 305 51(1) & 0.017 340 149(5)\\
$(0,1)$  & $\pm$ 1 & 0.002 687 162(4)    & 0.009 174 487 6(3) & 0.017 340 401(5)\\
\end{tabular}
\end{ruledtabular}
\end{table}
\endgroup

\begingroup
\squeezetable
\begin{table}[th]
\caption{Magic wavelengths expressed as photon energies (in a.u.) between the ground-state and
the first excied-state of HD$^+$ molecular ions.
The values correspond to the marked arrows in Figs.~\ref{magic1}--\ref{magic3}. The number in
parentheses represents the computational uncertainty.
}\label{magic-tab}
\begin{ruledtabular}
\begin{tabular}{lcllll}
\multicolumn{1}{l}{$M$} & \multicolumn{4}{c}{Magic-wavelengths}   \\
 \hline
 0       & 0.000 768 659 980(1) & 0.009 260 494 92(1) & 0.016 800 815 862(5) & 0.017 170 143 339(1) & 0.017 343 53(1)\\
$\pm 1$ & 0.017 189 146 9(2)  & 0.017 331 08(1)\\
\end{tabular}
\end{ruledtabular}
\end{table}
\endgroup

\subsection{Long-range interactions}

\begin{table}[th]
\caption{Long-range dispersion coefficients $C_6$,
$C_8$ and $C_{10}$ (in a.u.) for a ground state H$_2^+$, D$_2^+$, or HD$^+$ ion
interacting with a ground-state H, He, or Li atom. The numbers in parentheses represent
the computational uncertainties.}\label{tab6}
\begin{ruledtabular}
\begin{tabular}{llll}
\multicolumn{1}{l}{System} & \multicolumn{1}{c}{$C_6$}
& \multicolumn{1}{c}{$C_8$} & \multicolumn{1}{c}{$C_{10}$} \\
\hline
H$_2^+$-H           & 4.891 143 017 14(1)      & 90.316 962 31(1)     & 1807.210 076(2) \\
D$_2^+$-H           & 4.797 060 197 49(1)      & 87.850 021 22(1)     & 1756.323 945(2)\\
HD$^+$-H            & 5.381 569 069 96(1)      & 99.592 513 40(2)     & 2023.687 265(3)\\
H$_2^+$-He          & 2.195 917 825 1(1)       & 28.404 530 92(1)     & 368.784 69(1) \\
D$_2^+$-He          & 2.161 390 926 5(1)       & 27.641 661 03(1)     & 357.632 88(1)\\
HD$^+$-He           & 2.344 144 702 7(3)       & 31.043 628 96(3)     & 416.428 89(1)\\
H$_2^+$-Li          & 47.684(2)                & 2838.66(3)           & 168607(1) \\
D$_2^+$-Li          & 46.411(2)                & 2754.84(2)           & 163881(1)\\
HD$^+$-Li           & 66.498(2)                & 3354.24(2)           & 196257(1)\\
\end{tabular}
\end{ruledtabular}
\end{table}

Spectroscopic measurements of  the  Rydberg states of the
hydrogen molecules $\textrm{H}_2$ and $\textrm{D}_2$
have been performed by several groups~\cite{HerJun82,sturrus91a,DavGueSti90b,OstWueMer04}.
The data can be explained  in terms of the long-range polarization potential model,
in which, among other terms, the multipole polarizabilities of the parent
molecular ions $\textrm{H}_2^+$ or $\textrm{D}_2^+$
enter as parameters in the effective potentials of the multipole expansion
of the ion interaction with the distant charge~\cite{HerJun82,ChaPulEyl84,StuHesArc88}.
An elaborate polarization potential model was developed for analysis of experiments on
the highly-excited Rydberg states of the
hydrogen and deuterium molecules~\cite{Stu88,JunDabHer89,DavGueSti90a,sturrus91a,jacobson97a,jacobson00a}.
Its application yielded the experimental values for the static polarizabilities~\cite{jacobson00a} given in Table~\ref{tab2}.
Our nonadiabatic results for $\alpha_2 (0)$ and higher multipoles do not appear
to be readily applicable to
this particular model, which utilizes a separation of higher order
polarizabilities into electronic, vibrational, and rotational contributions.
For example, fits of the measured spectra utilize the electronic and
vibrational components of $\alpha_2(0)$; the rotational component
is treated as a higher order perturbation~\cite{Stu88,jacobson97a} and handled separately.

We used the dynamic multipole polarizabilities  to calculate the
long-range dispersion coefficients $C_6$, $C_8$, and $C_{10}$ for
the interaction between a ground state H, He, or Li atom and a
ground state $\textrm{H}_2^+$, $\textrm{D}_2^+$, or $\textrm{HD}^+$
ion. The results are given in Table~\ref{tab6}. The detailed
expressions for the coefficients were given in Refs.~\cite{tang09a}
and \cite{tang09b}. For the atoms we used methods described
previously. For H, the energies and matrix elements are obtained
using the Sturmian basis set to diagonalize the hydrogen
Hamiltonian~\cite{yan96a}, while for He and Li, the wave functions
are expanded as a linear combination of Hylleraas
functions~\cite{yan96a,tang09a}.

When the atom is in the ground state
but the molecular ion (denoted by ``b") is in an
excited $L_b$ state with magnetic quantum number $M_b$,  the
leading terms of the second-order interaction energy are
\begin{eqnarray}
V_{ab}=-\frac{C_6^{M_b}}{R^6}-\frac{C_8^{M_b}}{R^8}-\cdots,\label{eq:t56} .
\end{eqnarray}
The detailed expressions for  $C_6^{M_b}$ and $C_8^{M_b}$ are given in
Refs.~\cite{tang09a} and \cite{tang09b}.

\begin{table}[th]
\caption{ Long-range dispersion coefficients $C_6^{M_b}$
and $C_8^{M_b}$ (in a.u.) for an H$_2^+$, D$_2^+$, or HD$^+$
ion in the  first excited-state with magnetic quantum number $M_b$
interacting with a ground-state H, He, or Li atom.  The numbers
in parentheses represent the computational
uncertainties.}\label{tab7}
\begin{ruledtabular}
\begin{tabular}{llll}
\multicolumn{1}{l}{System} & \multicolumn{1}{c}{$M_b$}
& \multicolumn{1}{c}{$C_6^{M_b}$} & \multicolumn{1}{c}{$C_{8}^{M_b}$} \\
\hline
H$_2^+$-H          & 0        & 5.542 473 599 4(1)       & 114.730 417 8(1)  \\
D$_2^+$-H          & 0        & 5.417 791 267 5(1)       & 110.820 998 3(1)      \\
HD$^+$-H           & 0        & 6.233 633(2)             & 136.097 48(2)     \\
H$_2^+$-H          & $\pm 1$  & 4.581 282 889 3(1)       &  71.564 803 4(1) \\
D$_2^+$-H          & $\pm 1$  & 4.494 441 451 7(1)       &  69.756 367 1(2) \\
HD$^+$-H           & $\pm 1$  & 4.968 859(2)             &  74.802 77(3) \\
\hline
H$_2^+$-He         & 0        & 2.449 741 778 8(1)      &  36.967 531 92(1) \\
D$_2^+$-He         & 0        & 2.404 518 138 9(2)      &  35.706 083 48(2)  \\
HD$^+$-He          & 0        & 2.659 058(1)            &  43.666 737(2) \\
H$_2^+$-He         & $\pm 1$  & 2.075 030 168 4(1)      &  20.670 396 1(1) \\
D$_2^+$-He         & $\pm 1$  & 2.042 792 749 7(1)      &  20.159 006 1(1) \\
HD$^+$-He          & $\pm 1$  & 2.191 658(1)            &  21.291 70(4) \\
\hline
H$_2^+$-Li         & 0        & 55.361(1)             & 3500.16(2) \\
D$_2^+$-Li         & 0        & 53.635(1)             & 3377.72(2) \\
HD$^+$-Li          & 0        & 81.810(1)             & 4431.18(2) \\
H$_2^+$-Li         & $\pm 1$  & 44.045(1)             & 2466.08(1) \\
D$_2^+$-Li         & $\pm 1$  & 42.893(1)             & 2395.81(1) \\
HD$^+$-Li          & $\pm 1$  & 59.047(1)             & 2772.99(1) \\
\end{tabular}
\end{ruledtabular}
\end{table}

Table~\ref{tab7} lists the dispersion coefficients of  H$_2^+$, D$_2^+$, and HD$^+$ ions in the first
excited state  $(L_b=1)$ interacting with the ground-state H, He, and Li atoms.
As above, the atomic properties were taken from Ref.~\cite{yan96a}.
Note that the precision of
the calculated
$C_6^{M_b}$ and $C_8^{M_b}$ for the excited-state  HD$^+$ interacting with H and He atoms
is less than that for the H$_2^+$ and D$_2^+$ ions.
In the case of interactions with Li,  the accuracy of the coefficients is limited
by the accuracy of the Li calculations.

\section{Conclusion}\label {conclusion}

We calculated the static and dynamic multipole polarizabilities and hyerpolarizabilities for
the ground and first excited states  of
H$_2^+$, D$_2^+$, and HD$^+$ in the non-relativistic limit by using
correlated Hylleraas basis sets
without using the
Born-Oppenheimer approximation. For the static dipole polarizability of H$_2^+$, the
present value is the most accurate to date.  The hyperpolarizabilities were calculated
without derivatives (not using finite field methods) for
H$_2^+$ and its isotopomers.
The present high precision values can not only be taken as
a benchmark for testing other theoretical methods, but may also lay a foundation for investigating the relativistic and QED effects on
polarizabilites and hyperpolarizabilities and assist in planning experimental research
on hydrogen molecular ions.

\begin{acknowledgments}
We are grateful to Prof. J. Mitroy for comments and to Prof. W. G.
Sturrus for helpful correspondence. This work was supported by the
National Basic Research Program of China under Grant Nos.
2010CB832803 and 2012CB821305 and by NNSF of China under Grant Nos.
11104323, 11274348. Z.-C.Y. was supported by NSERC of Canada and by
the computing facilities of ACEnet and SHARCnet,  and in part by the
CAS/SAFEA International Partnership Program for Creative Research
Teams. ITAMP is supported in part by a grant from the NSF to the
Smithsonian Astrophysical Observatory and Harvard University.
\end{acknowledgments}


\begin{thebibliography}{59}
\expandafter\ifx\csname
natexlab\endcsname\relax\def\natexlab#1{#1}\fi
\expandafter\ifx\csname bibnamefont\endcsname\relax
  \def\bibnamefont#1{#1}\fi
\expandafter\ifx\csname bibfnamefont\endcsname\relax
  \def\bibfnamefont#1{#1}\fi
\expandafter\ifx\csname citenamefont\endcsname\relax
  \def\citenamefont#1{#1}\fi
\expandafter\ifx\csname url\endcsname\relax
  \def\url#1{\texttt{#1}}\fi
\expandafter\ifx\csname urlprefix\endcsname\relax\def\urlprefix{URL
}\fi \providecommand{\bibinfo}[2]{#2}
\providecommand{\eprint}[2][]{\url{#2}}

\bibitem[{\citenamefont{Yan et~al.}(2003)\citenamefont{Yan, Zhang, and
  Li}}]{yan03b}
\bibinfo{author}{\bibfnamefont{Z.-C.} \bibnamefont{Yan}},
  \bibinfo{author}{\bibfnamefont{J.-Y.} \bibnamefont{Zhang}}, \bibnamefont{and}
  \bibinfo{author}{\bibfnamefont{Y.}~\bibnamefont{Li}}, \bibinfo{journal}{Phys.
  Rev. A} \textbf{\bibinfo{volume}{67}}, \bibinfo{pages}{062504}
  (\bibinfo{year}{2003}).

\bibitem[{\citenamefont{{Zhang} and {Yan}}(2004)}]{zhang04a}
\bibinfo{author}{\bibfnamefont{J.-Y.} \bibnamefont{{Zhang}}} \bibnamefont{and}
  \bibinfo{author}{\bibfnamefont{Z.-C.} \bibnamefont{{Yan}}},
  \bibinfo{journal}{J.~Phys.~B} \textbf{\bibinfo{volume}{37}},
  \bibinfo{pages}{723} (\bibinfo{year}{2004}).

\bibitem[{\citenamefont{Bhatia and Drachman}(1999)}]{bhatia99a}
\bibinfo{author}{\bibfnamefont{A.~K.} \bibnamefont{Bhatia}} \bibnamefont{and}
  \bibinfo{author}{\bibfnamefont{R.~J.} \bibnamefont{Drachman}},
  \bibinfo{journal}{Phys. Rev. A} \textbf{\bibinfo{volume}{59}},
  \bibinfo{pages}{205} (\bibinfo{year}{1999}).

\bibitem[{\citenamefont{Bhatia and Drachman}(2000)}]{bhatia00a}
\bibinfo{author}{\bibfnamefont{A.~K.} \bibnamefont{Bhatia}} \bibnamefont{and}
  \bibinfo{author}{\bibfnamefont{R.~J.} \bibnamefont{Drachman}},
  \bibinfo{journal}{Phys. Rev. A} \textbf{\bibinfo{volume}{61}},
  \bibinfo{pages}{032503} (\bibinfo{year}{2000}).

\bibitem[{\citenamefont{Zhong et~al.}(2012)\citenamefont{Zhong, Zhang, Yan, and
  Shi}}]{ZhoZhaYan12}
\bibinfo{author}{\bibfnamefont{Z.-X.} \bibnamefont{Zhong}},
  \bibinfo{author}{\bibfnamefont{P.-P.} \bibnamefont{Zhang}},
  \bibinfo{author}{\bibfnamefont{Z.-C.} \bibnamefont{Yan}}, \bibnamefont{and}
  \bibinfo{author}{\bibfnamefont{T.-Y.} \bibnamefont{Shi}},
  \bibinfo{journal}{Phys. Rev. A} \textbf{\bibinfo{volume}{86}},
  \bibinfo{pages}{064502} (\bibinfo{year}{2012}).

\bibitem[{\citenamefont{Olivares~Pil\'on and Baye}(2013)}]{OliPilBay13}
\bibinfo{author}{\bibfnamefont{H.}~\bibnamefont{Olivares~Pil\'on}}
  \bibnamefont{and} \bibinfo{author}{\bibfnamefont{D.}~\bibnamefont{Baye}},
  \bibinfo{journal}{Phys. Rev. A} \textbf{\bibinfo{volume}{88}},
  \bibinfo{pages}{032502} (\bibinfo{year}{2013}).

\bibitem[{\citenamefont{Stanke and Adamowicz}(2013)}]{StaAda13}
\bibinfo{author}{\bibfnamefont{M.}~\bibnamefont{Stanke}} \bibnamefont{and}
  \bibinfo{author}{\bibfnamefont{L.}~\bibnamefont{Adamowicz}},
  \bibinfo{journal}{J. Phys. Chem. A} \textbf{\bibinfo{volume}{117}},
  \bibinfo{pages}{10129} (\bibinfo{year}{2013}).

\bibitem[{\citenamefont{Korobov and Zhong}(2012)}]{KorZho12}
\bibinfo{author}{\bibfnamefont{V.~I.} \bibnamefont{Korobov}} \bibnamefont{and}
  \bibinfo{author}{\bibfnamefont{Z.-X.} \bibnamefont{Zhong}},
  \bibinfo{journal}{Phys. Rev. A} \textbf{\bibinfo{volume}{86}},
  \bibinfo{pages}{044501} (\bibinfo{year}{2012}).

\bibitem[{\citenamefont{Kedziera et~al.}(2006)\citenamefont{Kedziera, Stanke,
  Bubin, Barysz, and Adamowicz}}]{KedStaBub06}
\bibinfo{author}{\bibfnamefont{D.}~\bibnamefont{Kedziera}},
  \bibinfo{author}{\bibfnamefont{M.}~\bibnamefont{Stanke}},
  \bibinfo{author}{\bibfnamefont{S.}~\bibnamefont{Bubin}},
  \bibinfo{author}{\bibfnamefont{M.}~\bibnamefont{Barysz}}, \bibnamefont{and}
  \bibinfo{author}{\bibfnamefont{L.}~\bibnamefont{Adamowicz}},
  \bibinfo{journal}{J. Chem. Phys.} \textbf{\bibinfo{volume}{125}},
  \bibinfo{eid}{084303} (\bibinfo{year}{2006}).

\bibitem[{\citenamefont{Bishop et~al.}(1980)\citenamefont{Bishop, Cheung, and
  Buckingham}}]{BisCheBuc80}
\bibinfo{author}{\bibfnamefont{D.}~\bibnamefont{Bishop}},
  \bibinfo{author}{\bibfnamefont{L.}~\bibnamefont{Cheung}}, \bibnamefont{and}
  \bibinfo{author}{\bibfnamefont{A.}~\bibnamefont{Buckingham}},
  \bibinfo{journal}{Molec. Phys.} \textbf{\bibinfo{volume}{41}},
  \bibinfo{pages}{1225} (\bibinfo{year}{1980}).

\bibitem[{\citenamefont{Pandey and Santry}(1980)}]{PanSan80}
\bibinfo{author}{\bibfnamefont{P.~K.~K.} \bibnamefont{Pandey}}
  \bibnamefont{and} \bibinfo{author}{\bibfnamefont{D.~P.}
  \bibnamefont{Santry}}, \bibinfo{journal}{J. Chem. Phys.}
  \textbf{\bibinfo{volume}{73}}, \bibinfo{pages}{2899} (\bibinfo{year}{1980}).

\bibitem[{\citenamefont{Adamowicz and Bartlett}(1986)}]{AdaBar86}
\bibinfo{author}{\bibfnamefont{L.}~\bibnamefont{Adamowicz}} \bibnamefont{and}
  \bibinfo{author}{\bibfnamefont{R.~J.} \bibnamefont{Bartlett}},
  \bibinfo{journal}{J. Chem. Phys.} \textbf{\bibinfo{volume}{84}},
  \bibinfo{pages}{4988} (\bibinfo{year}{1986}), \bibinfo{note}{;\textbf{86},
  7250E (1987)}.

\bibitem[{\citenamefont{Bishop}(1990)}]{Bis90}
\bibinfo{author}{\bibfnamefont{D.~M.} \bibnamefont{Bishop}},
  \bibinfo{journal}{Rev. Mod. Phys.} \textbf{\bibinfo{volume}{62}},
  \bibinfo{pages}{343} (\bibinfo{year}{1990}).

\bibitem[{\citenamefont{Bishop and Kirtman}(1992)}]{BisKir92}
\bibinfo{author}{\bibfnamefont{D.~M.} \bibnamefont{Bishop}} \bibnamefont{and}
  \bibinfo{author}{\bibfnamefont{B.}~\bibnamefont{Kirtman}},
  \bibinfo{journal}{J. Chem. Phys.} \textbf{\bibinfo{volume}{97}},
  \bibinfo{pages}{5255} (\bibinfo{year}{1992}).

\bibitem[{\citenamefont{Shelton and Rice}(1994)}]{SheRic94}
\bibinfo{author}{\bibfnamefont{D.~P.} \bibnamefont{Shelton}} \bibnamefont{and}
  \bibinfo{author}{\bibfnamefont{J.~E.} \bibnamefont{Rice}},
  \bibinfo{journal}{Chem. Rev.} \textbf{\bibinfo{volume}{94}},
  \bibinfo{pages}{3} (\bibinfo{year}{1994}).

\bibitem[{\citenamefont{Bishop}({1998})}]{Bis98}
\bibinfo{author}{\bibfnamefont{D.}~\bibnamefont{Bishop}}, in
  \emph{\bibinfo{booktitle}{{Advances in Chemical Physics}}}, edited by
  \bibinfo{editor}{\bibfnamefont{I.}~\bibnamefont{Prigogine}} \bibnamefont{and}
  \bibinfo{editor}{\bibfnamefont{S.~A.} \bibnamefont{Rice}}
  (\bibinfo{publisher}{Wiley}, \bibinfo{address}{New York},
  \bibinfo{year}{{1998}}), vol. \bibinfo{volume}{{104}} of
  \emph{\bibinfo{series}{{Advances in Chemical Physics}}}, pp.
  \bibinfo{pages}{{1--40}}.

\bibitem[{\citenamefont{Olivares~Pil\'on and Baye}(2012)}]{OliPilBay12}
\bibinfo{author}{\bibfnamefont{H.}~\bibnamefont{Olivares~Pil\'on}}
  \bibnamefont{and} \bibinfo{author}{\bibfnamefont{D.}~\bibnamefont{Baye}},
  \bibinfo{journal}{J. Phys. B} \textbf{\bibinfo{volume}{45}},
  \bibinfo{pages}{235101} (\bibinfo{year}{2012}).

\bibitem[{\citenamefont{Koelemeij}(2011)}]{koelemeij11a}
\bibinfo{author}{\bibfnamefont{J.}~\bibnamefont{Koelemeij}},
  \bibinfo{journal}{Phys. Chem. Chem. Phys.} \textbf{\bibinfo{volume}{13}},
  \bibinfo{pages}{18844} (\bibinfo{year}{2011}).

\bibitem[{\citenamefont{Cafiero et~al.}(2003)\citenamefont{Cafiero, Bubin, and
  Adamowicz}}]{CafBubAda03}
\bibinfo{author}{\bibfnamefont{M.}~\bibnamefont{Cafiero}},
  \bibinfo{author}{\bibfnamefont{S.}~\bibnamefont{Bubin}}, \bibnamefont{and}
  \bibinfo{author}{\bibfnamefont{L.}~\bibnamefont{Adamowicz}},
  \bibinfo{journal}{Phys. Chem. Chem. Phys.} \textbf{\bibinfo{volume}{5}},
  \bibinfo{pages}{1491} (\bibinfo{year}{2003}).

\bibitem[{\citenamefont{Ingamells et~al.}(1998)\citenamefont{Ingamells,
  Papadopoulos, Handy, and Willetts}}]{IngPapHan98}
\bibinfo{author}{\bibfnamefont{V.~E.} \bibnamefont{Ingamells}},
  \bibinfo{author}{\bibfnamefont{M.~G.} \bibnamefont{Papadopoulos}},
  \bibinfo{author}{\bibfnamefont{N.~C.} \bibnamefont{Handy}}, \bibnamefont{and}
  \bibinfo{author}{\bibfnamefont{A.}~\bibnamefont{Willetts}},
  \bibinfo{journal}{J. Chem. Phys.} \textbf{\bibinfo{volume}{109}},
  \bibinfo{pages}{1845} (\bibinfo{year}{1998}).

\bibitem[{\citenamefont{Schiller
  et~al.}(2014{\natexlab{a}})\citenamefont{Schiller, Bakalov, and
  Korobov}}]{SchBakKor14}
\bibinfo{author}{\bibfnamefont{S.}~\bibnamefont{Schiller}},
  \bibinfo{author}{\bibfnamefont{D.}~\bibnamefont{Bakalov}}, \bibnamefont{and}
  \bibinfo{author}{\bibfnamefont{V.~I.} \bibnamefont{Korobov}}
  (\bibinfo{year}{2014}{\natexlab{a}}), \eprint{arxiv.org:1402.1789}.

\bibitem[{\citenamefont{Karr et~al.}(2011)\citenamefont{Karr, Hilico, and
  Korobov}}]{KarHilKor11}
\bibinfo{author}{\bibfnamefont{J.-P.} \bibnamefont{Karr}},
  \bibinfo{author}{\bibfnamefont{L.}~\bibnamefont{Hilico}}, \bibnamefont{and}
  \bibinfo{author}{\bibfnamefont{V.~I.} \bibnamefont{Korobov}},
  \bibinfo{journal}{Can. J. Phys.} \textbf{\bibinfo{volume}{89}},
  \bibinfo{pages}{103} (\bibinfo{year}{2011}).

\bibitem[{\citenamefont{Shi et~al.}(2013)\citenamefont{Shi, Herskind, Drewsen,
  and Chuang}}]{ShiHerDre13}
\bibinfo{author}{\bibfnamefont{M.}~\bibnamefont{Shi}},
  \bibinfo{author}{\bibfnamefont{P.~F.} \bibnamefont{Herskind}},
  \bibinfo{author}{\bibfnamefont{M.}~\bibnamefont{Drewsen}}, \bibnamefont{and}
  \bibinfo{author}{\bibfnamefont{I.~L.} \bibnamefont{Chuang}},
  \bibinfo{journal}{New J. Phys.} \textbf{\bibinfo{volume}{15}},
  \bibinfo{pages}{113019} (\bibinfo{year}{2013}).

\bibitem[{\citenamefont{Bakalov and Schiller}(2012)}]{BakSch12}
\bibinfo{author}{\bibfnamefont{D.}~\bibnamefont{Bakalov}} \bibnamefont{and}
  \bibinfo{author}{\bibfnamefont{S.}~\bibnamefont{Schiller}},
  \bibinfo{journal}{Hyperfine Int.} \textbf{\bibinfo{volume}{210}},
  \bibinfo{pages}{25} (\bibinfo{year}{2012}).

\bibitem[{\citenamefont{Bishop and Solunac}(1985)}]{BisSol85}
\bibinfo{author}{\bibfnamefont{D.~M.} \bibnamefont{Bishop}} \bibnamefont{and}
  \bibinfo{author}{\bibfnamefont{S.~A.} \bibnamefont{Solunac}},
  \bibinfo{journal}{Phys. Rev. Lett.} \textbf{\bibinfo{volume}{55}},
  \bibinfo{pages}{1986} (\bibinfo{year}{1985}).

\bibitem[{\citenamefont{Sturrus et~al.}(1991)\citenamefont{Sturrus, Hessels,
  Arcuni, and Lundeen}}]{sturrus91a}
\bibinfo{author}{\bibfnamefont{W.~G.} \bibnamefont{Sturrus}},
  \bibinfo{author}{\bibfnamefont{E.~A.} \bibnamefont{Hessels}},
  \bibinfo{author}{\bibfnamefont{P.~W.} \bibnamefont{Arcuni}},
  \bibnamefont{and} \bibinfo{author}{\bibfnamefont{S.~R.}
  \bibnamefont{Lundeen}}, \bibinfo{journal}{Phys. Rev. A}
  \textbf{\bibinfo{volume}{44}}, \bibinfo{pages}{3032} (\bibinfo{year}{1991}).

\bibitem[{\citenamefont{Jacobson et~al.}(1997)\citenamefont{Jacobson, Fisher,
  Fehrenbach, Sturrus, and Lundeen}}]{jacobson97a}
\bibinfo{author}{\bibfnamefont{P.~L.} \bibnamefont{Jacobson}},
  \bibinfo{author}{\bibfnamefont{D.~S.} \bibnamefont{Fisher}},
  \bibinfo{author}{\bibfnamefont{C.~W.} \bibnamefont{Fehrenbach}},
  \bibinfo{author}{\bibfnamefont{W.~G.} \bibnamefont{Sturrus}},
  \bibnamefont{and} \bibinfo{author}{\bibfnamefont{S.~R.}
  \bibnamefont{Lundeen}}, \bibinfo{journal}{Phys. Rev. A}
  \textbf{\bibinfo{volume}{56}}, \bibinfo{pages}{R4361} (\bibinfo{year}{1997}).

\bibitem[{\citenamefont{Jacobson et~al.}(2000)\citenamefont{Jacobson, Komara,
  Sturrus, and Lundeen}}]{jacobson00a}
\bibinfo{author}{\bibfnamefont{P.~L.} \bibnamefont{Jacobson}},
  \bibinfo{author}{\bibfnamefont{R.~A.} \bibnamefont{Komara}},
  \bibinfo{author}{\bibfnamefont{W.~G.} \bibnamefont{Sturrus}},
  \bibnamefont{and} \bibinfo{author}{\bibfnamefont{S.~R.}
  \bibnamefont{Lundeen}}, \bibinfo{journal}{Phys. Rev. A}
  \textbf{\bibinfo{volume}{62}}, \bibinfo{pages}{012509}
  (\bibinfo{year}{2000}).

\bibitem[{\citenamefont{{Mohr} et~al.}(2008)\citenamefont{{Mohr}, {Taylor}, and
  {Newell}}}]{mohr08a}
\bibinfo{author}{\bibfnamefont{P.~J.} \bibnamefont{{Mohr}}},
  \bibinfo{author}{\bibfnamefont{B.~N.} \bibnamefont{{Taylor}}},
  \bibnamefont{and} \bibinfo{author}{\bibfnamefont{D.~B.}
  \bibnamefont{{Newell}}}, \bibinfo{journal}{J.~Phys.~Chem.~Ref.~Data}
  \textbf{\bibinfo{volume}{37}}, \bibinfo{pages}{1187} (\bibinfo{year}{2008}).

\bibitem{CODATA10a}
After our calculations were started the
2010 CODATA values were released for which the masses of the proton and deuteron
differ in their last digits from the corresponding 2006 CODATA values; the changes do not significantly affect our results.

\bibitem[{\citenamefont{Bishop et~al.}(1988)\citenamefont{Bishop, Lam, and
  Epstein}}]{BisLamEps88}
\bibinfo{author}{\bibfnamefont{D.~M.} \bibnamefont{Bishop}},
  \bibinfo{author}{\bibfnamefont{B.}~\bibnamefont{Lam}}, \bibnamefont{and}
  \bibinfo{author}{\bibfnamefont{S.~T.} \bibnamefont{Epstein}},
  \bibinfo{journal}{J. Chem. Phys.} \textbf{\bibinfo{volume}{88}},
  \bibinfo{pages}{337} (\bibinfo{year}{1988}).

\bibitem[{\citenamefont{Schiller
  et~al.}(2014{\natexlab{b}})\citenamefont{Schiller, Bakalov, Bekbaev, and
  Korobov}}]{SchBakKor14b}
\bibinfo{author}{\bibfnamefont{S.}~\bibnamefont{Schiller}},
  \bibinfo{author}{\bibfnamefont{D.}~\bibnamefont{Bakalov}},
  \bibinfo{author}{\bibfnamefont{A.~K.} \bibnamefont{Bekbaev}},
  \bibnamefont{and} \bibinfo{author}{\bibfnamefont{V.~I.}
  \bibnamefont{Korobov}},
  \bibinfo{journal}{Phys. Rev. A } \textbf{\bibinfo{volume}{89}},
  \bibinfo{pages}{052521} (\bibinfo{year}{2014}).

\bibitem[{\citenamefont{Yan and Drake}(1996)}]{YanDra96}
\bibinfo{author}{\bibfnamefont{Z.-C.} \bibnamefont{Yan}} \bibnamefont{and}
  \bibinfo{author}{\bibfnamefont{G.}~\bibnamefont{Drake}},
  \bibinfo{journal}{Chem. Phys. Lett.} \textbf{\bibinfo{volume}{259}},
  \bibinfo{pages}{96} (\bibinfo{year}{1996}).

\bibitem[{\citenamefont{Tang et~al.}(2010)\citenamefont{Tang, Yan, Shi, and
  Mitroy}}]{tang10a}
\bibinfo{author}{\bibfnamefont{L.-Y.} \bibnamefont{Tang}},
  \bibinfo{author}{\bibfnamefont{Z.-C.} \bibnamefont{Yan}},
  \bibinfo{author}{\bibfnamefont{T.-Y.} \bibnamefont{Shi}}, \bibnamefont{and}
  \bibinfo{author}{\bibfnamefont{J.}~\bibnamefont{Mitroy}},
  \bibinfo{journal}{Phys. Rev. A} \textbf{\bibinfo{volume}{81}},
  \bibinfo{pages}{042521} (\bibinfo{year}{2010}).

\bibitem[{\citenamefont{{Tang} et~al.}(2009{\natexlab{a}})\citenamefont{{Tang},
  {Yan}, {Shi}, and {Babb}}}]{tang09a}
\bibinfo{author}{\bibfnamefont{L.-Y.} \bibnamefont{{Tang}}},
  \bibinfo{author}{\bibfnamefont{Z.-C.} \bibnamefont{{Yan}}},
  \bibinfo{author}{\bibfnamefont{T.-Y.} \bibnamefont{{Shi}}}, \bibnamefont{and}
  \bibinfo{author}{\bibfnamefont{J.~F.} \bibnamefont{{Babb}}},
  \bibinfo{journal}{Phys.~Rev.~A} \textbf{\bibinfo{volume}{79}},
  \bibinfo{pages}{062712} (\bibinfo{year}{2009}{\natexlab{a}}).

\bibitem[{\citenamefont{{Tang} et~al.}(2009{\natexlab{b}})\citenamefont{{Tang},
  {Zhang}, {Yan}, {Shi}, {Babb}, and {Mitroy}}}]{tang09b}
\bibinfo{author}{\bibfnamefont{L.-Y.} \bibnamefont{{Tang}}},
  \bibinfo{author}{\bibfnamefont{J.-Y.} \bibnamefont{{Zhang}}},
  \bibinfo{author}{\bibfnamefont{Z.-C.} \bibnamefont{{Yan}}},
  \bibinfo{author}{\bibfnamefont{T.-Y.} \bibnamefont{{Shi}}},
  \bibinfo{author}{\bibfnamefont{J.~F.} \bibnamefont{{Babb}}},
  \bibnamefont{and} \bibinfo{author}{\bibfnamefont{J.}~\bibnamefont{{Mitroy}}},
  \bibinfo{journal}{Phys.~Rev.~A} \textbf{\bibinfo{volume}{80}},
  \bibinfo{pages}{042511} (\bibinfo{year}{2009}{\natexlab{b}}).

\bibitem[{\citenamefont{{Pipin} and {Bishop}}(1992)}]{pipin92a}
\bibinfo{author}{\bibfnamefont{J.}~\bibnamefont{{Pipin}}} \bibnamefont{and}
  \bibinfo{author}{\bibfnamefont{D.~M.} \bibnamefont{{Bishop}}},
  \bibinfo{journal}{J.~Phys.~B} \textbf{\bibinfo{volume}{25}},
  \bibinfo{pages}{17} (\bibinfo{year}{1992}).

\bibitem[{\citenamefont{Korobov}(2006)}]{korobov06a}
\bibinfo{author}{\bibfnamefont{V.~I.} \bibnamefont{Korobov}},
  \bibinfo{journal}{Phys. Rev. A} \textbf{\bibinfo{volume}{74}},
  \bibinfo{pages}{052506} (\bibinfo{year}{2006}).

\bibitem[{\citenamefont{{Mohr} and {Taylor}}(2005)}]{mohr05a}
\bibinfo{author}{\bibfnamefont{P.~J.} \bibnamefont{{Mohr}}} \bibnamefont{and}
  \bibinfo{author}{\bibfnamefont{B.~N.} \bibnamefont{{Taylor}}},
  \bibinfo{journal}{Rev.~Mod.~Phys.} \textbf{\bibinfo{volume}{77}},
  \bibinfo{pages}{1} (\bibinfo{year}{2005}).

\bibitem[{\citenamefont{Tian et~al.}(2012)\citenamefont{Tian, Tang, Zhong, Yan,
  and Shi}}]{TiaTanZho12}
\bibinfo{author}{\bibfnamefont{Q.-L.} \bibnamefont{Tian}},
  \bibinfo{author}{\bibfnamefont{L.-Y.} \bibnamefont{Tang}},
  \bibinfo{author}{\bibfnamefont{Z.-X.} \bibnamefont{Zhong}},
  \bibinfo{author}{\bibfnamefont{Z.-C.} \bibnamefont{Yan}}, \bibnamefont{and}
  \bibinfo{author}{\bibfnamefont{T.-Y.} \bibnamefont{Shi}},
  \bibinfo{journal}{J. Chem. Phys.} \textbf{\bibinfo{volume}{137}},
  \bibinfo{pages}{024311} (\bibinfo{year}{2012}).

\bibitem[{\citenamefont{Korobov}(2008)}]{Kor08}
\bibinfo{author}{\bibfnamefont{V.~I.} \bibnamefont{Korobov}},
  \bibinfo{journal}{Phys. Rev. A} \textbf{\bibinfo{volume}{77}},
  \bibinfo{pages}{022509} (\bibinfo{year}{2008}).

\bibitem[{\citenamefont{Zhong et~al.}(2009)\citenamefont{Zhong, Yan, and
  Shi}}]{ZhoYanShi09}
\bibinfo{author}{\bibfnamefont{Z.-X.} \bibnamefont{Zhong}},
  \bibinfo{author}{\bibfnamefont{Z.-C.} \bibnamefont{Yan}}, \bibnamefont{and}
  \bibinfo{author}{\bibfnamefont{T.-Y.} \bibnamefont{Shi}},
  \bibinfo{journal}{Phys. Rev. A} \textbf{\bibinfo{volume}{79}},
  \bibinfo{pages}{064502} (\bibinfo{year}{2009}).

\bibitem[{\citenamefont{{Hilico} et~al.}(2001)\citenamefont{{Hilico}, {Billy},
  {Gr{\'e}maud}, and {Delande}}}]{hilico01a}
\bibinfo{author}{\bibfnamefont{L.}~\bibnamefont{{Hilico}}},
  \bibinfo{author}{\bibfnamefont{N.}~\bibnamefont{{Billy}}},
  \bibinfo{author}{\bibfnamefont{B.}~\bibnamefont{{Gr{\'e}maud}}},
  \bibnamefont{and}
  \bibinfo{author}{\bibfnamefont{D.}~\bibnamefont{{Delande}}},
  \bibinfo{journal}{J. Phys. B} \textbf{\bibinfo{volume}{34}},
  \bibinfo{pages}{491} (\bibinfo{year}{2001}).

\bibitem[{\citenamefont{Korobov}(2001)}]{korobov01b}
\bibinfo{author}{\bibfnamefont{V.~I.} \bibnamefont{Korobov}},
  \bibinfo{journal}{Phys. Rev. A} \textbf{\bibinfo{volume}{63}},
  \bibinfo{pages}{044501} (\bibinfo{year}{2001}).

\bibitem[{\citenamefont{{Moss} and {Valenzano}}(2002)}]{moss02a}
\bibinfo{author}{\bibfnamefont{R.~E.} \bibnamefont{{Moss}}} \bibnamefont{and}
  \bibinfo{author}{\bibfnamefont{L.}~\bibnamefont{{Valenzano}}},
  \bibinfo{journal}{Mol. Phys.} \textbf{\bibinfo{volume}{100}},
  \bibinfo{pages}{649} (\bibinfo{year}{2002}).

\bibitem[{\citenamefont{Moss}(1999)}]{Mos99b}
\bibinfo{author}{\bibfnamefont{R.~E.} \bibnamefont{Moss}}, \bibinfo{journal}{J.
  Phys. B} \textbf{\bibinfo{volume}{32}}, \bibinfo{pages}{L89}
  (\bibinfo{year}{1999}).

\bibitem[{\citenamefont{Taylor et~al.}(1999)\citenamefont{Taylor, Dalgarno, and
  Babb}}]{taylor99a}
\bibinfo{author}{\bibfnamefont{J.~M.} \bibnamefont{Taylor}},
  \bibinfo{author}{\bibfnamefont{A.}~\bibnamefont{Dalgarno}}, \bibnamefont{and}
  \bibinfo{author}{\bibfnamefont{J.~F.} \bibnamefont{Babb}},
  \bibinfo{journal}{Phys. Rev. A} \textbf{\bibinfo{volume}{60}},
  \bibinfo{pages}{R2630} (\bibinfo{year}{1999}).

\bibitem[{\citenamefont{Bishop and Lam}(1988)}]{BisLam88}
\bibinfo{author}{\bibfnamefont{D.~M.} \bibnamefont{Bishop}} \bibnamefont{and}
  \bibinfo{author}{\bibfnamefont{B.}~\bibnamefont{Lam}}, \bibinfo{journal}{Mol.
  Phys.} \textbf{\bibinfo{volume}{65}}, \bibinfo{pages}{679}
  (\bibinfo{year}{1988}).

\bibitem[{\citenamefont{Orr and Ward}(1971)}]{OrrWar71}
\bibinfo{author}{\bibfnamefont{B.}~\bibnamefont{Orr}} \bibnamefont{and}
  \bibinfo{author}{\bibfnamefont{J.}~\bibnamefont{Ward}},
  \bibinfo{journal}{Mol. Phys.} \textbf{\bibinfo{volume}{20}},
  \bibinfo{pages}{513} (\bibinfo{year}{1971}).

\bibitem[{\citenamefont{Bishop and Lam}(1987)}]{BisLam87}
\bibinfo{author}{\bibfnamefont{D.~M.} \bibnamefont{Bishop}} \bibnamefont{and}
  \bibinfo{author}{\bibfnamefont{B.}~\bibnamefont{Lam}}, \bibinfo{journal}{Mol.
  Phys.} \textbf{\bibinfo{volume}{62}}, \bibinfo{pages}{721}
  (\bibinfo{year}{1987}).

\bibitem[{\citenamefont{LeBlanc and Thywissen}(2007)}]{LeBThy07}
\bibinfo{author}{\bibfnamefont{L.~J.} \bibnamefont{LeBlanc}} \bibnamefont{and}
  \bibinfo{author}{\bibfnamefont{J.~H.} \bibnamefont{Thywissen}},
  \bibinfo{journal}{Phys. Rev. A} \textbf{\bibinfo{volume}{75}},
  \bibinfo{pages}{053612} (\bibinfo{year}{2007}).

\bibitem[{\citenamefont{Herzberg and Jungen}(1982)}]{HerJun82}
\bibinfo{author}{\bibfnamefont{G.}~\bibnamefont{Herzberg}} \bibnamefont{and}
  \bibinfo{author}{\bibfnamefont{C.}~\bibnamefont{Jungen}},
  \bibinfo{journal}{J. Chem. Phys.} \textbf{\bibinfo{volume}{77}},
  \bibinfo{pages}{5876} (\bibinfo{year}{1982}).

\bibitem[{\citenamefont{{Davies}
  et~al.}(1990{\natexlab{a}})\citenamefont{{Davies}, {Guest}, and
  {Stickland}}}]{DavGueSti90b}
\bibinfo{author}{\bibfnamefont{P.~B.} \bibnamefont{{Davies}}},
  \bibinfo{author}{\bibfnamefont{M.~A.} \bibnamefont{{Guest}}},
  \bibnamefont{and} \bibinfo{author}{\bibfnamefont{R.~J.}
  \bibnamefont{{Stickland}}}, \bibinfo{journal}{J. Chem. Phys.}
  \textbf{\bibinfo{volume}{93}}, \bibinfo{pages}{5417}
  (\bibinfo{year}{1990}{\natexlab{a}}).

\bibitem[{\citenamefont{{Osterwalder} et~al.}(2004)\citenamefont{{Osterwalder},
  {W{\"u}est}, {Merkt}, and {Jungen}}}]{OstWueMer04}
\bibinfo{author}{\bibfnamefont{A.}~\bibnamefont{{Osterwalder}}},
  \bibinfo{author}{\bibfnamefont{A.}~\bibnamefont{{W{\"u}est}}},
  \bibinfo{author}{\bibfnamefont{F.}~\bibnamefont{{Merkt}}}, \bibnamefont{and}
  \bibinfo{author}{\bibfnamefont{C.}~\bibnamefont{{Jungen}}},
  \bibinfo{journal}{J. Chem. Phys.} \textbf{\bibinfo{volume}{121}},
  \bibinfo{pages}{11810} (\bibinfo{year}{2004}).

\bibitem[{\citenamefont{{Chang} et~al.}(1984)\citenamefont{{Chang},
  {Pulchtopek}, and {Eyler}}}]{ChaPulEyl84}
\bibinfo{author}{\bibfnamefont{E.~S.} \bibnamefont{{Chang}}},
  \bibinfo{author}{\bibfnamefont{S.}~\bibnamefont{{Pulchtopek}}},
  \bibnamefont{and} \bibinfo{author}{\bibfnamefont{E.~E.}
  \bibnamefont{{Eyler}}}, \bibinfo{journal}{J. Chem. Phys.}
  \textbf{\bibinfo{volume}{80}}, \bibinfo{pages}{601} (\bibinfo{year}{1984}).

\bibitem[{\citenamefont{{Sturrus} et~al.}(1988)\citenamefont{{Sturrus},
  {Hessels}, {Arcuni}, and {Lundeen}}}]{StuHesArc88}
\bibinfo{author}{\bibfnamefont{W.~G.} \bibnamefont{{Sturrus}}},
  \bibinfo{author}{\bibfnamefont{E.~A.} \bibnamefont{{Hessels}}},
  \bibinfo{author}{\bibfnamefont{P.~W.} \bibnamefont{{Arcuni}}},
  \bibnamefont{and} \bibinfo{author}{\bibfnamefont{S.~R.}
  \bibnamefont{{Lundeen}}}, \bibinfo{journal}{Phys. Rev. A}
  \textbf{\bibinfo{volume}{38}}, \bibinfo{pages}{135} (\bibinfo{year}{1988}).

\bibitem[{\citenamefont{Sturrus}(1988)}]{Stu88}
\bibinfo{author}{\bibfnamefont{W.~G.} \bibnamefont{Sturrus}}, Ph.D. thesis,
  \bibinfo{school}{Notre Dame University} (\bibinfo{year}{1988}).

\bibitem[{\citenamefont{{Jungen} et~al.}(1989)\citenamefont{{Jungen},
  {Dabrowski}, {Herzberg}, and {Kendall}}}]{JunDabHer89}
\bibinfo{author}{\bibfnamefont{C.}~\bibnamefont{{Jungen}}},
  \bibinfo{author}{\bibfnamefont{I.}~\bibnamefont{{Dabrowski}}},
  \bibinfo{author}{\bibfnamefont{G.}~\bibnamefont{{Herzberg}}},
  \bibnamefont{and} \bibinfo{author}{\bibfnamefont{D.~J.~W.}
  \bibnamefont{{Kendall}}}, \bibinfo{journal}{J. Chem. Phys.}
  \textbf{\bibinfo{volume}{91}}, \bibinfo{pages}{3926} (\bibinfo{year}{1989}).

\bibitem[{\citenamefont{{Davies}
  et~al.}(1990{\natexlab{b}})\citenamefont{{Davies}, {Guest}, and
  {Stickland}}}]{DavGueSti90a}
\bibinfo{author}{\bibfnamefont{P.~B.} \bibnamefont{{Davies}}},
  \bibinfo{author}{\bibfnamefont{M.~A.} \bibnamefont{{Guest}}},
  \bibnamefont{and} \bibinfo{author}{\bibfnamefont{R.~J.}
  \bibnamefont{{Stickland}}}, \bibinfo{journal}{J. Chem. Phys.}
  \textbf{\bibinfo{volume}{93}}, \bibinfo{pages}{5408}
  (\bibinfo{year}{1990}{\natexlab{b}}).

\bibitem[{\citenamefont{Yan et~al.}(1996)\citenamefont{Yan, Babb, Dalgarno, and
  Drake}}]{yan96a}
\bibinfo{author}{\bibfnamefont{Z.~C.} \bibnamefont{Yan}},
  \bibinfo{author}{\bibfnamefont{J.~F.} \bibnamefont{Babb}},
  \bibinfo{author}{\bibfnamefont{A.}~\bibnamefont{Dalgarno}}, \bibnamefont{and}
  \bibinfo{author}{\bibfnamefont{G.~W.~F.} \bibnamefont{Drake}},
  \bibinfo{journal}{Phys.~Rev.~A} \textbf{\bibinfo{volume}{54}},
  \bibinfo{pages}{2824} (\bibinfo{year}{1996}).

\end{thebibliography}

\end{document}